\documentclass[journal]{IEEEtran}
\usepackage[figuresright]{rotating}
\usepackage{amssymb}
\usepackage{amsthm}
\usepackage{multicol}
\usepackage{subfigure}
\usepackage{graphicx}
\usepackage{graphics}
\usepackage{epstopdf}
\usepackage{fullpage}
\usepackage{latexsym,amsmath}
\usepackage{stmaryrd}
\usepackage{algorithm,algorithmic}
\usepackage{subfigure}
\usepackage{caption}
\usepackage{amsfonts}
\usepackage{xcolor,multirow}
\usepackage{cite}
\usepackage{bm}
\usepackage{url}
\usepackage{diagbox}
\usepackage{array}
\usepackage{booktabs}
\usepackage{footmisc} 
\usepackage{pifont}

\newtheorem{theorem}{Theorem}
\newtheorem{definition}{Definition}
\newtheorem{remark}{Remark}

\ifCLASSINFOpdf
\else
\fi
\begin{document}
	
\title{Quaternion Tensor Train Rank Minimization with Sparse Regularization
	in a Transformed Domain for Quaternion Tensor Completion}

\author{Jifei~Miao,\,\, Kit~Ian~Kou,\,\,Liqiao~Yang,\,\,Dong~Cheng 

\thanks{Jifei~Miao is with the School of Mathematics and Statistics, Yunnan University, Kunming, Yunnan, 650091, China (e-mail: jifmiao@163.com)}
\thanks{Kit~Ian~Kou is with the Department of Mathematics, Faculty of Science
	and Technology, University of Macau, Macau 999078, China  (e-mail:
	kikou@umac.mo)}
\thanks{Liqiao~Yang is with the Department of Mathematics, Faculty of Science
	and Technology, University of Macau, Macau 999078, China  (e-mail:
	liqiaoyoung@163.com)}
\thanks{Dong~Cheng is with the Department of Mathematics, Faculty of Arts and Sciences, Beijing Normal University, Zhuhai 519087, China(e-mail:
	chengdong720@163.com)}
}


\maketitle

\begin{abstract}
The tensor train rank (TT-rank) has achieved promising results in tensor completion due to its ability to capture the global low-rankness of higher-order ($>3$) tensors. On the other hand, recently, quaternions have proven to be a very suitable framework for encoding color pixels, and have obtained outstanding performance in various color image processing tasks. In this paper, the quaternion tensor train (QTT) decomposition is presented, and based on that the quaternion TT-rank (QTT-rank) is naturally defined, which are the generalizations of their counterparts in the real number field. In addition, to utilize the local sparse prior of the quaternion tensor, a general and flexible transform framework is defined. Combining both the global low-rank and local sparse priors of the quaternion tensor, we propose a novel quaternion tensor completion model, \emph{i.e.}, QTT-rank minimization with sparse regularization
in a transformed domain. Specifically, we use the quaternion weighted nuclear norm (QWNN) of mode-$n$ canonical unfolding quaternion matrices to characterize the global low-QTT-rankness, and the $l_{1}$-norm of the quaternion tensor in a transformed domain to characterize the local sparse property. Moreover, to enable the QTT-rank minimization to handle color images and better handle color videos, we generalize KA, a tensor augmentation method, to quaternion tensors and define quaternion KA (QKA), which is a helpful pretreatment step for QTT-rank based optimization problems. The numerical experiments on color images and color videos inpainting tasks indicate the advantages of the proposed method over the state-of-the-art ones.
\end{abstract}
\begin{IEEEkeywords}
Quaternion tensor train decomposition, quaternion tensor completion, sparse regularization, color image inpainting, color video.
\end{IEEEkeywords}

\IEEEpeerreviewmaketitle

\section{Introduction}\label{sec1}
\IEEEPARstart{T}ensor provides a natural way to represent multidimensional data. For example, a color image is a third-order tensor, a color video is a fourth-order tensor with an additional index in the time variable. In recent years, low-rank tensor completion (LRTC)-based techniques have made a great success in the application of image and video inpainting \cite{DBLP:journals/sigpro/LongLCZ19,DBLP:journals/tsp/ZhangA17,DBLP:journals/tip/BenguaPTD17}. Generally, the LRTC problem is modeled as
\begin{equation}\label{equint1}\small
	\begin{split}
	&\mathop{{\rm{min}}}\limits_{\mathcal{X}}\ {\rm{rank}}(\mathcal{X})\\ 
	&\ \text{s.t.}\ P_{\Omega}(\mathcal{X})=P_{\Omega}(\mathcal{T}),
\end{split}	
\end{equation}
where $\mathcal{X}\in \mathbb{R}^{I_{1}\times I_{2}\times \ldots \times I_{N}}$  is a completed output $N$-th order tensor,  $\mathcal{T}\in \mathbb{R}^{I_{1}\times I_{2}\times \ldots \times I_{N}}$ is the observed $N$-th order tensor, and $P_{\Omega}(\cdot)$ is the projection
operator on $\Omega$ which is the index of observed entries. There are different types of definitions of tensor ranks to characterize the low-rankness of tensors, such as CP rank \cite{DBLP:journals/siamrev/KoldaB09}, Tucker rank \cite{DBLP:conf/icdm/KoldaS08}, tubal rank \cite{DBLP:journals/pami/LuFCLLY20}, tensor train rank (TT-rank) \cite{DBLP:journals/siamsc/Oseledets11}, \emph{etc.} Among these definitions of tensor ranks, the TT-rank recently has shown a powerful capacity for characterizing the correlations between different modes in higher-order tensors, which as a result has achieved great success in LRTC. 	Given an $N$-th order tensor $\mathcal{X}\in \mathbb{R}^{I_{1}\times I_{2}\times \ldots \times I_{N}}$, the TT-rank \cite{DBLP:conf/bigdata/BenguaPT15} is defined as
\begin{equation}\label{equttrank}\small
	{\rm{rank_{TT}}}(\mathcal{X})=({\rm{rank}}(\mathbf{X}_{[1]}), \ldots,{\rm{rank}}(\mathbf{X}_{[N-1]})),
\end{equation}
where $\mathbf{X}_{[n]}\in \mathbb{R}^{\Pi_{j=1}^{n}I_{j}\times \Pi_{j=n+1}^{N}I_{j}}$ denotes the Mode-$n$ canonical unfolding of
$\mathcal{X}$ \cite{DBLP:journals/siamsc/Oseledets11}. Due to the TT decomposition being free from the ``curse of dimensionality'' \cite{DBLP:journals/siamsc/Oseledets11} and the TT-rank can capture the
global correlation of higher-order tensors, researchers have developed a series of methods for TT-rank minimization. For LRTC problem, the authors in \cite{DBLP:journals/tip/BenguaPTD17} proposed two TT-rank minimization models. The first one named SiLRTC-TT minimizes the weighted sum of the convex nuclear norm surrogate of TT-rank, \emph{i.e.},
\begin{equation}\label{equint2}\small
	\begin{split}
		&\mathop{{\rm{min}}}\limits_{\mathcal{X}}\ \sum_{k=1}^{N-1}\alpha_{k}\|\mathbf{X}_{[k]}\|_{\ast}\\ 
		&\ \text{s.t.}\ P_{\Omega}(\mathcal{X})=P_{\Omega}(\mathcal{T}),
	\end{split}
\end{equation}
where $\alpha_{k}$ for $k=1,2,\ldots,N-1$  are positive parameters. Another one named TMac-TT uses matrix factorization to approximate the TT-rank, \emph{i.e.},
\begin{equation}\label{equint3}\small
	\begin{split}
		&\mathop{{\rm{min}}}\limits_{\mathbf{P}_{k},\mathbf{Q}_{k},\mathcal{X}}\ \sum_{k=1}^{N-1}\frac{\alpha_{k}}{2}\|\mathbf{P}_{k}\mathbf{Q}_{k}-\mathbf{X}_{[k]}\|_{F}^{2}\\ 
		&\ \text{s.t.}\ P_{\Omega}(\mathcal{X})=P_{\Omega}(\mathcal{T}),
	\end{split}
\end{equation}
where $\mathbf{P}_{k}\in \mathbb{R}^{\Pi_{j=1}^{n}I_{j}\times r_{k}}$ and $\mathbf{Q}_{k}\in \mathbb{R}^{r_{k}\times \Pi_{j=n+1}^{N}I_{j}}$ are factor matrices, and $r_{k}$ denotes the rank
of $\mathbf{X}_{[k]}$. For TMac-TT, the estimation of $r_{k}$ is a challenging task, which directly affects the algorithm performance \cite{DBLP:journals/tip/BenguaPTD17}.

On the other hand, quaternion, as an elegant color image representation tool, has attracted much attention in the field of color image processing \cite{DBLP:journals/ijon/YuZY19,DBLP:journals/tip/ZouKW16,DBLP:journals/tip/ChenXZ20,DBLP:journals/nla/JiaNS19}. The quaternion processes a color image holistically as a vector field and handles the coupling between the color channels naturally, which allows the color information of the source image is fully used. Consequently, plenty of low-rank quaternion matrix completion (LRQMC) methods have been proposed in recent years and shown promising results in color image inpainting tasks \cite{DBLP:journals/tip/ChenXZ20,DBLP:journals/tip/MiaoK22,DBLP:journals/isci/YangMK22}. Furthermore, as a generalization of LRTC in the quaternion domain, we first proposed low-rank quaternion tensor completion (LRQTC) model in our previous work \cite{DBLP:journals/pr/MiaoKL20}, \emph{i.e.},
\begin{equation}\small
	\label{equint4}
	\begin{split}
		&\mathop{{\rm{min}}}\limits_{\dot{\mathcal{X}}}\quad \sum_{k=1}^{N}\alpha_{k}\|\dot{\mathbf{X}}_{(n)}\|_{\ast}\\
		&\ \text{s.t.}\ P_{\Omega}(\dot{\mathcal{X}})=P_{\Omega}(\dot{\mathcal{T}}),
	\end{split}
\end{equation}
where $\dot{\mathcal{X}}\in \mathbb{H}^{I_{1}\times I_{2}\times \ldots \times I_{N}}$  is a completed output $N$-th order quaternion tensor,  $\dot{\mathcal{T}}\in \mathbb{H}^{I_{1}\times I_{2}\times \ldots \times I_{N}}$ is the observed $N$-th order quaternion tensor, and  $\dot{\mathbf{X}}_{(n)}\in \mathbb{H}^{I_{n}\times \Pi_{j=1,j\neq n}^{N}I_{j}}$ denotes the Mode-$n$ unfolding of
$\dot{\mathcal{X}}$ \cite{DBLP:journals/pr/MiaoKL20}. The model (\ref{equint4}) is directly following the definition of Tucker rank, \emph{i.e.},
\begin{equation}\label{equint5}\small
	{\rm{rank_{tucker}}}(\mathcal{X})=({\rm{rank}}(\mathbf{X}_{(1)}), \ldots,{\rm{rank}}(\mathbf{X}_{(N)})).
\end{equation}
However, a conceptual flaw of Tucker rank is that its components are the ranks of matrices constructed based on an unbalanced unfolding scheme (one mode versus the rest). The upper bound for each individual rank is usually small and may not be suitable for describing the global information of a tensor \cite{DBLP:journals/tip/BenguaPTD17}. Based on this consideration, we define the
quaternion tensor train (QTT) decomposition of quaternion tensors and the corresponding QTT-rank in this paper. The details can be founded in Section \ref{sec3}.

In addition, the low-rank component always indicates that the data in practice also have intrinsically sparse property \cite{DBLP:journals/pieee/WrightMMSHY10}. One possible approach is to exploit the sparse information of the complete tensor in some domains, such as the transform domain where many signals have an inherently sparse structure \cite{DBLP:journals/nla/WangLC21,DBLP:conf/cvpr/YangWHM08}. Motivated by this, in this paper, to utilize the local sparse prior of the quaternion tensor, a general and flexible transform framework is defined. Under the framework, any appropriate quaternion multidimensional discrete transform can be contained. The details can be founded in Section \ref{sec4_sub1}.

Moreover, to enable the QTT-rank minimization to handle color images and better handle
color videos, we generalize KA \cite{DBLP:journals/tip/BenguaPTD17}, a tensor augmentation method, to quaternion tensors and define quaternion KA (QKA), which is a helpful pretreatment step for QTT-rank
based optimization problems. The details can be founded in Section \ref{sec_QKA}. To summarize, the main contributions of this paper are listed as follows:
\begin{itemize}
	\item The QTT decomposition is first presented, and based on it the QTT-rank is naturally defined, which are the generalizations of their counterparts in the real number field.
	\item A general and flexible transform framework for quaternion tensors is defined, which can reveal the essential sparsity of visual data.
	\item Combining both the global low-QTT-rank and local sparse priors of the quaternion tensor, we propose a novel LRQTC model.  Then, the model is optimized by applying the quaternion-based alternating direction method of
	multipliers (ADMM) algorithm. 
	\item  The QKA is defined for QTT-rank minimization approaches to handle color images and better handle color videos. Extensive experiments are conducted to demonstrate the effectiveness and superiority of the proposed method for inpainting problems of color images and color videos.
\end{itemize}

The remainder of this paper is organized as follows. Section \ref{sec2} introduces some
main notations and needed basic knowledge of quaternions. Section \ref{sec3} presents the QTT decomposition and the QTT-rank. Section \ref{sec4} gives the proposed LRQCT model and the solving algorithm. A quaternion tensor augmentation technique, 
\emph{i.e.}, QKA is introduced in Section \ref{sec_QKA}. Section \ref{sec_6}
provides extensive experiments to illustrate the performance of the proposed method. Finally, some conclusions are drawn in Section \ref{sec_7}.

\section{Preliminary}\label{sec2}
In this section, we list some main notations and introduce some needed basic knowledge of quaternions.
\subsection{Notations}
In this paper, $\mathbb{R}$, $\mathbb{C}$, and $\mathbb{H}$ respectively denote the real space, complex space, and quaternion space. A scalar, a vector, a matrix, and a tensor are written as $a$, $\mathbf{a}$, $\mathbf{A}$, and $\mathcal{A}$ respectively. $\dot{a}$,  $\dot{\mathbf{a}}$, $\dot{\mathbf{A}}$, and $\dot{\mathcal{A}}$ 
respectively represent a quaternion scalar, a quaternion vector, a quaternion matrix, and a quaternion tensor.  $(\cdot)^{\ast}$, $(\cdot)^{-1}$, and $(\cdot)^{H}$ denote the conjugation,  inverse, and  conjugate transpose, respectively.  $|\cdot|$, $\|\cdot\|_{F}$, $\|\cdot\|_{*}$, $\|\cdot\|_{w,*}$, and $\|\cdot\|_{1}$ are respectively the modulus, Frobenius norm, nuclear norm, weighted nuclear norm, and $l_{1}$-norm.
$\mathbf{X}_{(n)}$ and $\mathbf{X}_{[n]}$  respectively denote mode-$n$ unfolding and mode-$n$ canonical unfolding of quaternion tensor $\dot{\mathcal{X}}$. $\times_{n}$, $\otimes$, and $\langle \cdot, \cdot \rangle$ are respectively the $n$-mode product, the Kronecker product, and the inner product.


\subsection{Basic Knowledge of Quaternions}
Preliminaries of quaternions (including quaternion matrices and quaternion tensors) can be seen in Appendix \ref{a_qsec1}.

\section{QTT Decomposition and QTT-rank}\label{sec3}
As a generalization of the real-valued TT decomposition in \cite{DBLP:journals/siamsc/Oseledets11}, we define the following QTT decomposition: 
\begin{definition}(QTT Decomposition)
\label{qtt}
Given an $N$-th order quaternion tensor $\dot{\mathcal{X}}\in \mathbb{H}^{I_{1}\times I_{2}\times \ldots \times I_{N}}$, the QTT decomposition models each entry of $\mathcal{X}$ by a series
of quaternion matrix products as
\begin{equation}\label{eqqtt}\small
\dot{x}_{i_{1},i_{2},\ldots i_{N}}=\dot{\mathcal{G}}_{1}(:,i_{1},:)\dot{\mathcal{G}}_{2}(:,i_{2},:)\ldots\dot{\mathcal{G}}_{N}(:,i_{N},:),
\end{equation}
where $\dot{\mathcal{G}}_{n}\in \mathbb{H}^{r_{n-1}\times I_{n}\times r_{n}}, n=1,2\ldots,N$ with $r_{0}=r_{N}=1$ are third-order quaternion tensors called the \textbf{cores} of $\dot{\mathcal{X}}$ and
$(r_{1},r_{2},\ldots, r_{N-1})$ is a vector closely related to the QTT-rank defined below. The graphical of QTT
decomposition can be seen in Figure \ref{qtt_fig}.
\begin{figure*}[htbp]
	\centering
	\includegraphics[width=14cm,height=2.8cm]{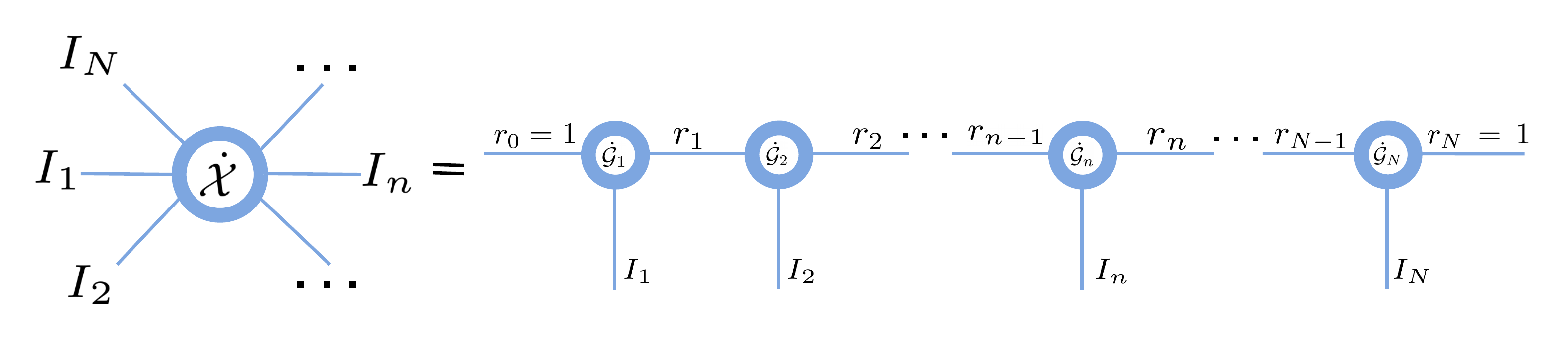} 
	\caption{The graphical of QTT decomposition.}
	\label{qtt_fig}
\end{figure*}
\end{definition}
\begin{remark}
In the index form the QTT decomposition (\ref{eqqtt}) can be written as
\begin{equation}\label{eqqtt2}\small
\begin{split}	
\dot{x}_{i_{1},i_{2},\ldots i_{N}}=\sum_{l_{1},l_{2},\ldots,l_{N-1}=1}^{r_{1},r_{2},\ldots,r_{N-1}}&\dot{\mathcal{G}}_{1}(1,i_{1},l_{1})\dot{\mathcal{G}}_{2}(l_{1},i_{2},l_{2})\ldots\\
&\dot{\mathcal{G}}_{N}(l_{N-1},i_{N},1).
\end{split}	
\end{equation}
In the following, we show that the QTT decomposition of a quaternion tensor $\dot{\mathcal{X}}\in \mathbb{H}^{I_{1}\times I_{2}\times \ldots \times I_{N}}$ can be derived recursively.

Firstly, treating the multi-index $(i_{2},\ldots i_{N})$ as a single index, then we have a following usual quaternion matrix decomposition:
\begin{equation}\label{equde1}\small
\dot{x}_{i_{1},i_{2},\ldots i_{N}}=\sum_{l_{1}=1}^{r_{1}}\dot{\mathcal{G}}_{1}(1,i_{1},l_{1})	\dot{\mathcal{H}}_{1}(l_{1},i_{2},\ldots i_{N}).	
\end{equation}
Next, for quaternion tensor $\dot{\mathcal{H}}_{1}$, the indices $(l_{1},i_{2})$ are separated from the rest, we also have a following usual quaternion matrix decomposition:
\begin{equation}\label{equde2}\small
\dot{\mathcal{H}}_{1}(l_{1},i_{2},\ldots i_{N})=\sum_{l_{2}=1}^{r_{2}}\dot{\mathcal{G}}_{2}(l_{1},i_{2},l_{2})	\dot{\mathcal{H}}_{2}(l_{2},i_{3},\ldots i_{N}).	
\end{equation}
Thus,
\begin{equation*}\small
\begin{split}
&\dot{x}_{i_{1},i_{2},\ldots i_{N}}\\
&=\sum_{l_{1},l_{2}=1}^{r_{1},r_{2}}\dot{\mathcal{G}}_{1}(1,i_{1},l_{1})	\dot{\mathcal{G}}_{2}(l_{1},i_{2},l_{2})\dot{\mathcal{H}}_{2}(l_{2},i_{3},\ldots i_{N}).	
\end{split}		
\end{equation*}
Proceeding in the above way, after $N$ steps, one can obtain a decomposition of the form (\ref{eqqtt2}).
\end{remark}

\begin{remark}
From (\ref{eqqtt}), one can easily find that the QTT decomposition of a quaternion tensor $\dot{\mathcal{X}}\in \mathbb{H}^{I_{1}\times I_{2}\times \ldots \times I_{N}}$ is not unique: we can insert the identity $\dot{\mathbf{Q}}\dot{\mathbf{Q}}^{-1}$ between any two quaternion matrices in the series to obtain another decomposition. 
\end{remark}

\begin{remark}
Denote $I=\max I_{n}$ and $r=\max r_{n}$, the number of parameters in the QTT decomposition (\ref{eqqtt}) is upper bounded by $NIr^{2}$. To constitute a great
reduction compared to storing $\Pi_{i=1}^{N}I_{i}$ entries in $\dot{\mathcal{X}}$ explicitly, we tend to choose the $r_{1},r_{2},\ldots, r_{N-1}$ as small as possible. 
\end{remark}
In the following, we first define another matricization operation for a quaternion tensor $\dot{\mathcal{X}}\in \mathbb{H}^{I_{1}\times I_{2}\times \ldots \times I_{N}}$ and then give an important theorem. They are the direct source of the QTT-rank definition.
\begin{definition}(Mode-$n$ canonical unfolding)
	\label{Mcun}
	Mode-$n$ canonical unfolding of an $N$-th order quaternion tensor $\dot{\mathcal{X}}\in \mathbb{H}^{I_{1}\times I_{2}\times \ldots \times I_{N}}$ is defined as
	\begin{equation}\label{equMcun}\small
		\dot{\mathbf{X}}_{[n]}\in \mathbb{H}^{\Pi_{j=1}^{n}I_{j}\times \Pi_{j=n+1}^{N}I_{j}},	
	\end{equation}
within each dimension of the matrix $\mathbf{X}_{[n]}$, the indices are ordered lexicographically, which is the same as calling Matlab's ``reshape'' function on a multidimensional array with the specified target dimension, i.e., 
\begin{equation*}\small
	\begin{split}
	\dot{\mathbf{X}}_{[n]}&={\rm{reshape}}(\dot{\mathcal{X}},\Pi_{j=1}^{n}I_{j},\Pi_{j=n+1}^{N}I_{j})\\
	&:={\rm{unfold}}_{[n]}(\dot{\mathcal{X}}).
    \end{split}	
\end{equation*}
Conversely, the mode-$n$ canonical quaternion matrix $\mathbf{X}_{[n]}$ can be transformed back to the quaternion tensor $\dot{\mathcal{X}}$ by 
\begin{equation*}\small
	\begin{split}
\dot{\mathcal{X}}&={\rm{reshape}}(\dot{\mathbf{X}}_{[n]},I_{1}, I_{2} \ldots, I_{N})\\
	&:={\rm{fold}}_{[n]}(\dot{\mathbf{X}}_{[n]}).
	\end{split}
\end{equation*}
\end{definition}
Compared with the Mode-$n$ unfolding operation in the Definition \ref{nuft}, the Mode-$n$ canonical unfolding is a well-balanced matricization characterizing the correlation between the first $n$ and the rest $N-n$ dimensions of $\dot{\mathcal{X}}$. 

\begin{theorem}\label{theqtt}
For an $N$-th order quaternion tensor $\dot{\mathcal{X}}\in \mathbb{H}^{I_{1}\times I_{2}\times \ldots \times I_{N}}$ with a QTT decomposition (\ref{eqqtt}), it necessarily holds that
\begin{equation}\label{equin}\small
r_{n}\geq {\rm{rank}}(\dot{\mathbf{X}}_{[n]}),\quad n=1,2,\ldots,N-1,
\end{equation}
and it is furthermore possible to obtain a QTT decomposition such that equality holds.
\end{theorem}
The proof of Theorem \ref{theqtt} can be founded in  Appendix \ref{a_sec1}. As described in Remark $3$ that we tend to choose the $r_{1},r_{2},\ldots, r_{N-1}$ as small as possible. Theorem \ref{theqtt} tells us that ${\rm{rank}}(\dot{\mathbf{X}}_{[n]})$ is the lower bound of
$r_{n}$ and furthermore ${\rm{rank}}(\dot{\mathbf{X}}_{[n]})$ is an admissible choice for $r_{n}$ with $n=1,2,\ldots,N-1$.
Therefore, ${\rm{rank}}(\dot{\mathbf{X}}_{[1]}), {\rm{rank}}(\dot{\mathbf{X}}_{[2]}),\ldots,{\rm{rank}}(\dot{\mathbf{X}}_{[N-1]})$ are of interest to us, for which we give the following definition of QTT-rank.

\begin{definition}(QTT-rank)
	\label{qttrank}
	QTT-rank of an $N$-th order quaternion tensor $\dot{\mathcal{X}}\in \mathbb{H}^{I_{1}\times I_{2}\times \ldots \times I_{N}}$ is defined as
	\begin{equation}\label{equqttrank}\small
	{\rm{rank_{QTT}}}(\dot{\mathcal{X}})=({\rm{rank}}(\dot{\mathbf{X}}_{[1]}), \ldots,{\rm{rank}}(\dot{\mathbf{X}}_{[N-1]})).
	\end{equation}
\end{definition}
One can see that ${\rm{rank}}(\dot{\mathbf{X}}_{[n]})(n=1,2,\ldots,N-1)$ contains correlations between permutations of all modes of $\dot{\mathcal{X}}$, which gives an effective way to capture the global correlations of a quaternion tensor.

\section{Low-Rank Quaternion Tensor Completion Model And Solving Algorithm}\label{sec4}
\subsection{Sparse Regularization Method in a Transformed Domain}\label{sec4_sub1}
\begin{figure*}[htbp]
	\centering
	\subfigure[]{
		\includegraphics[width=13cm,height=3.5cm]{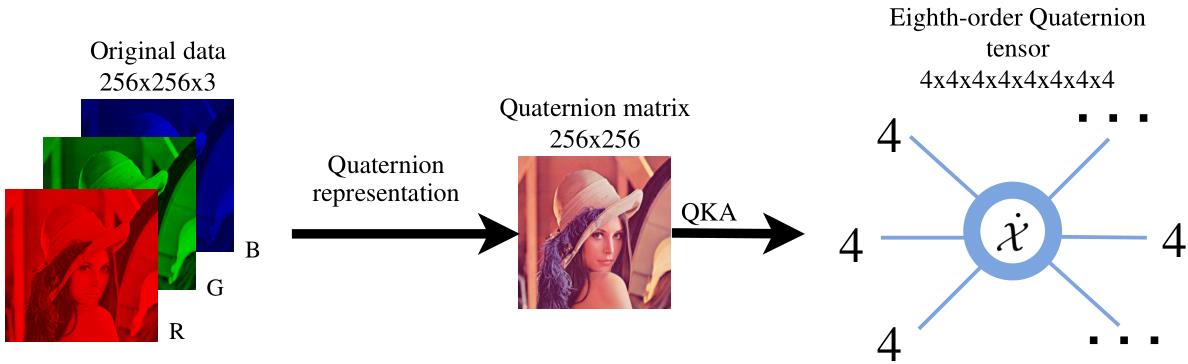}
	}\\
	\subfigure[]{
		\includegraphics[width=15.5cm,height=2cm]{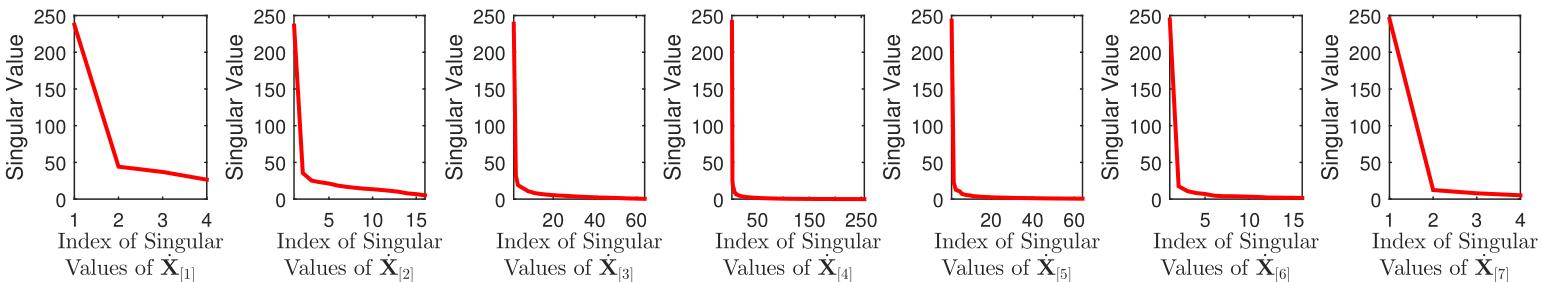}
	}\\
	\subfigure[]{
		\includegraphics[width=15.1cm,height=2.5cm]{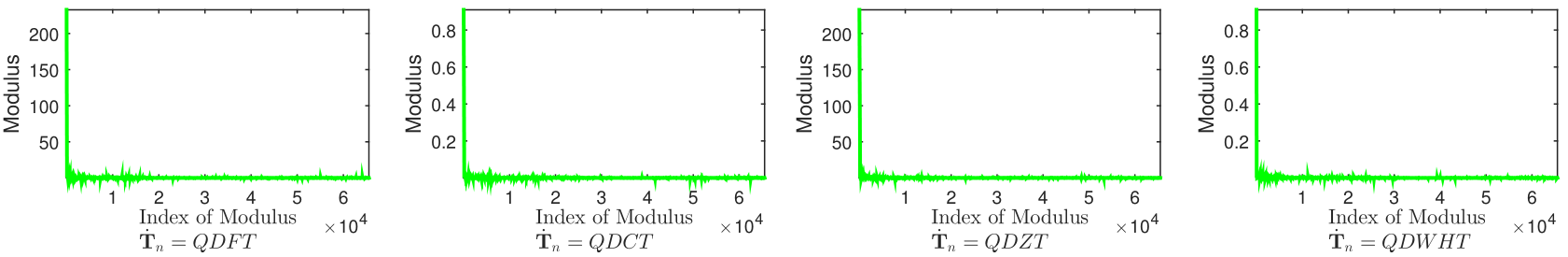}
	}
	\caption{(a) shows the process of a color image to higher-order quaternion tensor; (b) shows the low-QTT-rankness of the quaternion tensor generated by QKA; (c) shows the sparsity of the quaternion tensor in four transformed domains.}
	\label{example1}
\end{figure*}

In the following, we define a general and flexible transform leading the quaternion tensor to a sparse space.
\begin{definition}
The transform $\mathfrak{T}$ for an $N$-th order quaternion tensor $\dot{\mathcal{X}}\in \mathbb{H}^{I_{1}\times I_{2}\times \ldots \times I_{N}}$ is defined as
\begin{equation}\label{transform}\small
\mathfrak{T}(\dot{\mathcal{X}}):=\dot{\hat{\mathcal{X}}}=\dot{\mathcal{X}}\times_{1}\dot{\mathbf{T}}_{1}\times_{2}\dot{\mathbf{T}}_{2}\times\ldots\times_{N}\dot{\mathbf{T}}_{N},
\end{equation}
where $\dot{\mathbf{T}}_{n}\, (n=1,2,\ldots,N)$ are unitary quaternion transform matrices with appropriate size, e.g., quaternion discrete Fourier transforms (QDFT), quaternion discrete Cosine transforms (QDCT), and any other appropriate transforms. The inverse transform of $\mathfrak{T}$ is directly defined as
\begin{equation}\label{itransform}\small
	\mathfrak{T}^{-1}(\hat{\mathcal{X}}):=\dot{\mathcal{X}}=\hat{\mathcal{X}}\times_{1}\dot{\mathbf{T}}_{1}^{-1}\times_{2}\dot{\mathbf{T}}_{2}^{-1}\times\ldots\times_{N}\dot{\mathbf{T}}_{N}^{-1}.
\end{equation}
\end{definition} 
\begin{remark}
	The defined $\mathfrak{T}(\dot{\mathcal{X}})$ is a general and flexible framework. Firstly,  $\dot{\mathbf{T}}_{n}\, (n=1,2,\ldots,N)$ in (\ref{transform}) can be any appropriate transforms including but not limited to QDFT, QDCT, quaternion discrete Z transforms (QDZT), quaternion discrete Walsh-Hadamard transform (QDWHT). Furthermore, in some specific applications, $\dot{\mathbf{T}}_{n}\, (n=1,2,\ldots,N)$ in (\ref{transform}) can be a mixture of different types of transforms. Figure \ref{example1}(c) gives several examples of the defined transform $\mathfrak{T}$ on a given quaternion tensor generated by the color image ``lena''. One can find that for different kinds of transforms the defined transform $\mathfrak{T}$ do lead to a sparse space (the modulus of the vast majority of entries of the quaternion tensor in  transformed domains are very close to zero).
\end{remark} 

\begin{remark}
	The above mentioned QDFT, QDCT, QDZT, and QDWHT are quaternion versions of their traditional counterparts, i.e., DFT \cite{duhamel1990fast}, DCT \cite{jain1989fundamentals}, DZT \cite{rabiner1975theory}, and DWHT \cite{fino1976unified}. In Appendix \ref{a_sec2}, we give a general and simple way to transform these traditional transforms into quaternion versions by using the Cayley-Dickson form of the quaternion matrix.
\end{remark} 

\subsection{Proposed Model}
Based on the defined QTT-rank and the sparse regularization method in a transformed domain, we propose the following LRQTC model:
\begin{equation}\small
	\label{equmodel1}
	\begin{split}
		&\mathop{{\rm{min}}}\limits_{\dot{\mathcal{X}}}\ {\rm{rank_{QTT}}}(\dot{\mathcal{X}})+ \lambda\|\mathfrak{T}(\dot{\mathcal{X}})\|_{1}\\ 
		&\ \text{s.t.}\ P_{\Omega}(\dot{\mathcal{X}})=P_{\Omega}(\dot{\mathcal{T}}),
	\end{split}
\end{equation}
where $\lambda$ is a nonnegative parameter, $\dot{\mathcal{X}}\in \mathbb{H}^{I_{1}\times I_{2}\times \ldots \times I_{N}}$  is a completed output $N$-th order quaternion tensor,  $\dot{\mathcal{T}}\in \mathbb{H}^{I_{1}\times I_{2}\times \ldots \times I_{N}}$ is the observed $N$-th order quaternion tensor, and $P_{\Omega}(\cdot)$ is the projection
operator on $\Omega$ which is the index of observed entries. In order for (\ref{equmodel1}) to be solved, we first use an effective non-convex surrogate (QWNN) to replace the QTT-rank function, then we use the variable-splitting technique and introduce auxiliary quaternion variables $\dot{\mathcal{M}}_{k}\in \mathbb{H}^{I_{1}\times I_{2}\times \ldots \times I_{N}}$ and $\dot{\mathcal{E}}_{k}\in \mathbb{H}^{I_{1}\times I_{2}\times \ldots \times I_{N}},\, (k=1,2,\ldots, N-1)$ in (\ref{equmodel1}). Consequently, (\ref{equmodel1}) is finally transformed into the following solvable model:
\begin{equation}\small
	\label{equmodel2}
	\begin{split}
		&\mathop{{\rm{min}}}\limits_{\dot{\mathcal{X}},\dot{\mathcal{M}}_{k},\dot{\mathcal{E}}_{k}}\ \sum_{k=1}^{N-1}\alpha_{k}\|\dot{\mathcal{M}}_{k[k]}\|_{w,\ast}+ \lambda_{k}\|\dot{\mathcal{E}}_{k}\|_{1}\\ 
		&\quad\text{s.t.}\ P_{\Omega}(\dot{\mathcal{X}})=P_{\Omega}(\dot{\mathcal{T}})\\
		&\qquad\ \dot{\mathcal{X}}=\dot{\mathcal{M}}_{k},\ k=1,2,\ldots,N-1,\\
		&\qquad\ \mathfrak{T}(\dot{\mathcal{X}})=\dot{\mathcal{E}}_{k},\ k=1,2,\ldots,N-1,		
	\end{split}
\end{equation}
where $\alpha_{k}$ and $\lambda_{k}$ for $k=1,2,\ldots,N-1$  are nonnegative parameters.

\subsection{Solving Algorithm}
Based on the ADMM framework in the quaternion domain \cite{DBLP:journals/tsp/MiaoK20}, the augmented Lagrangian function of (\ref{equmodel2}) is defined as
\begin{equation}\label{equqlf}\small
\begin{split}
	&\!\!\!\!\!\!\mathcal{L}_{\mu}(\dot{\mathcal{X}},\{\dot{\mathcal{M}}_{k}\}_{k=1}^{N-1},\{\dot{\mathcal{E}}_{k}\}_{k=1}^{N-1},\{\dot{\mathcal{Y}}_{1k}\}_{k=1}^{N-1},\{\dot{\mathcal{Y}}_{2k}\}_{k=1}^{N-1})\\
	=&\sum_{k=1}^{N-1}\alpha_{k}\|\dot{\mathcal{M}}_{k[k]}\|_{w,\ast}+ \lambda_{k}\|\dot{\mathcal{E}}_{k}\|_{1}\\
	&+\mathfrak{R}(\langle\dot{\mathcal{Y}}_{1k},\dot{\mathcal{X}}-\dot{\mathcal{M}}_{k}\rangle)+\frac{\mu}{2}\|\dot{\mathcal{X}}-\dot{\mathcal{M}}_{k}\|_{F}^{2} \\ 
	&+\mathfrak{R}(\langle\dot{\mathcal{Y}}_{2k},\mathfrak{T}(\dot{\mathcal{X}})-\dot{\mathcal{E}}_{k}\rangle)+\frac{\mu}{2}\|\mathfrak{T}(\dot{\mathcal{X}})-\dot{\mathcal{E}}_{k}\|_{F}^{2}, 
\end{split}	
\end{equation}
where $\dot{\mathcal{Y}}_{1k}\in \mathbb{H}^{I_{1}\times I_{2}\times \ldots \times I_{N}}$ and $\dot{\mathcal{Y}}_{2k}\in \mathbb{H}^{I_{1}\times I_{2}\times \ldots \times I_{N}}$ for $k=1,2,\ldots,N-1$ are Lagrange multipliers, $\mu>0$ is the penalty parameter. Then, we use the following iterative scheme to solve the problem (\ref{equqlf}):
\begin{equation}\label{allsubp}\small\
	\left\{
	\begin{array}{lc}
		\dot{\mathcal{X}}^{(\tau+1)}=\mathop{{\rm{arg\, min}}}\limits_{P_{\Omega}(\dot{\mathcal{X}})=P_{\Omega}(\dot{\mathcal{T}})}\:\mathcal{L}_{\mu^{(\tau)}}\left(\dot{\mathcal{X}},\{\dot{\mathcal{M}}_{k}^{(\tau)}\}_{k=1}^{N-1},\right.\\
		\qquad\qquad\left.\{\dot{\mathcal{E}}_{k}^{(\tau)}\}_{k=1}^{N-1},
		 \{\dot{\mathcal{Y}}_{1k}^{(\tau)}\}_{k=1}^{N-1},\{\dot{\mathcal{Y}}_{2k}^{(\tau)}\}_{k=1}^{N-1}\right),	\\	
		\{\dot{\mathcal{M}}_{k}^{(\tau+1)}\}_{k=1}^{N-1}=\mathop{{\rm{arg\, min}}}\limits_{\{\dot{\mathcal{M}}_{k}\}_{k=1}^{N-1}}\:\mathcal{L}_{\mu^{(\tau)}}\left(\dot{\mathcal{X}}^{(\tau+1)},\{\dot{\mathcal{M}}_{k}\}_{k=1}^{N-1},\right. \\
		\qquad\qquad\qquad\quad\left. \{\dot{\mathcal{E}}_{k}^{(\tau)}\}_{k=1}^{N-1},  \{\dot{\mathcal{Y}}_{1k}^{(\tau)}\}_{k=1}^{N-1},\{\dot{\mathcal{Y}}_{2k}^{(\tau)}\}_{k=1}^{N-1}\right),\\	
		\{\dot{\mathcal{E}}_{k}^{(\tau+1)}\}_{k=1}^{N-1}=\mathop{{\rm{arg\, min}}}\limits_{\{\dot{\mathcal{E}}_{k}\}_{k=1}^{N-1}}\:\mathcal{L}_{\mu^{(\tau)}}\!\!\left(\!\dot{\mathcal{X}}^{(\tau+1)},\!\{\dot{\mathcal{M}}_{k}^{(\tau+1)}\}_{k=1}^{N-1},\right. \\
		\qquad\qquad\qquad\ \left. \{\dot{\mathcal{E}}_{k}\}_{k=1}^{N-1},  \{\dot{\mathcal{Y}}_{1k}^{(\tau)}\}_{k=1}^{N-1},\{\dot{\mathcal{Y}}_{2k}^{(\tau)}\}_{k=1}^{N-1}\right),\\	
		\{\dot{\mathcal{Y}}_{1k}^{(\tau+1)}\}_{k=1}^{N-1}=\{\dot{\mathcal{Y}}_{1k}^{(\tau)}\}_{k=1}^{N-1}+\mu^{(\tau)}\left(\dot{\mathcal{X}}^{(\tau+1)}-\right.\\
		\left.\qquad\qquad\qquad\ \{\dot{\mathcal{M}}_{k}^{(\tau+1)}\}_{k=1}^{N-1}\right),\\		\{\dot{\mathcal{Y}}_{2k}^{(\tau+1)}\}_{k=1}^{N-1}=\{\dot{\mathcal{Y}}_{2k}^{(\tau)}\}_{k=1}^{N-1}+\mu^{(\tau)}\left(\mathfrak{T}(\dot{\mathcal{X}}^{(\tau+1)})-\right.\\
		\left.\qquad\qquad\qquad\ \{\dot{\mathcal{E}}_{k}^{(\tau+1)}\}_{k=1}^{N-1}\right),
	\end{array}
	\right.
\end{equation}
where $\tau$ is the iteration index.
Next, we provide the details for solving the subproblems in (\ref{allsubp}).

\textbf{Updating $\dot{\mathcal{X}}$}:  In the $(\tau+1)$-th iteration, fixing the other variables at their latest values, $\dot{\mathcal{X}}^{(\tau+1)}$ is the optimal solution of the following problem:
\begin{equation}\small
	\begin{split}
		\dot{\mathcal{X}}^{(\tau+1)}=&\mathop{{\rm{arg\, min}}}\limits_{P_{\Omega}(\dot{\mathcal{X}})=P_{\Omega}(\dot{\mathcal{T}})}\ \sum_{k=1}^{N-1}\mathfrak{R}(\langle\dot{\mathcal{Y}}_{1k}^{(\tau)},\dot{\mathcal{X}}-\dot{\mathcal{M}}_{k}^{(\tau)}\rangle)\\
		&+\frac{\mu^{(\tau)}}{2}\|\dot{\mathcal{X}}-\dot{\mathcal{M}}_{k}^{(\tau)}\|_{F}^{2} \!+\!\mathfrak{R}(\langle\dot{\mathcal{Y}}_{2k}^{(\tau)},\mathfrak{T}(\dot{\mathcal{X}})\!-\!\dot{\mathcal{E}}_{k}^{(\tau)}\rangle)\\
		&+\frac{\mu^{(\tau)}}{2}\|\mathfrak{T}(\dot{\mathcal{X}})-\dot{\mathcal{E}}_{k}^{(\tau)}\|_{F}^{2}\\
		=&\mathop{{\rm{arg\, min}}}\limits_{P_{\Omega}(\dot{\mathcal{X}})=P_{\Omega}(\dot{\mathcal{T}})}\ \!\!\sum_{k=1}^{N-1}\frac{\mu^{(\tau)}}{2}\|\dot{\mathcal{X}}\!-\!\dot{\mathcal{M}}_{k}^{(\tau)}+\!\frac{\dot{\mathcal{Y}}_{1k}^{(\tau)}}{\mu^{(\tau)}}\|_{F}^{2}\\
		&+\frac{\mu^{(\tau)}}{2}\|\mathfrak{T}(\dot{\mathcal{X}})-\dot{\mathcal{E}}_{k}^{(\tau)}-\frac{\dot{\mathcal{Y}}_{2k}^{(\tau)}}{\mu^{(\tau)}}\|_{F}^{2}.
	\end{split}
\end{equation}
Since $\dot{\mathbf{T}}_{n}\, (n=1,2,\ldots,N)$ are unitary quaternion transform matrices in (\ref{transform}), we have
$\|\mathfrak{T}(\dot{\mathcal{X}})-\dot{\mathcal{E}}_{k}^{(\tau)}-\frac{\dot{\mathcal{Y}}_{2k}^{(\tau)}}{\mu^{(\tau)}}\|_{F}^{2}
=\|\dot{\mathcal{X}}-\mathfrak{T}^{-1}(\dot{\mathcal{E}}_{k}^{(\tau)}+\frac{\dot{\mathcal{Y}}_{2k}^{(\tau)}}{\mu^{(\tau)}})\|_{F}^{2}$, for $k=1,2,\ldots,N-1$. Hence, we can solve $\dot{\mathcal{X}}$ through
\begin{equation}\label{equupx}\small
	\begin{split}
		\dot{\mathcal{X}}^{(\tau+1)}
		=&\mathop{{\rm{arg\, min}}}\limits_{P_{\Omega}(\dot{\mathcal{X}})=P_{\Omega}(\dot{\mathcal{T}})}\ \!\sum_{k=1}^{N-1}\mu^{(\tau)}\|\dot{\mathcal{X}}-\frac{1}{2}\big(\dot{\mathcal{M}}_{k}^{(\tau)}-\!\frac{\dot{\mathcal{Y}}_{1k}^{(\tau)}}{\mu^{(\tau)}}\\
		&+\mathfrak{T}^{-1}(\dot{\mathcal{E}}_{k}^{(\tau)}+\frac{\dot{\mathcal{Y}}_{2k}^{(\tau)}}{\mu^{(\tau)}})\big)\|_{F}^{2}\\
		=&P_{\Omega^{c}}\!\!\left(\!\frac{\sum_{k=1}^{N-1}\big(\dot{\mathcal{M}}_{k}^{(\tau)}\!-\!\frac{\dot{\mathcal{Y}}_{1k}^{(\tau)}}{\mu^{(\tau)}}\!+\!\mathfrak{T}^{-1}(\dot{\mathcal{E}}_{k}^{(\tau)}\!+\!\frac{\dot{\mathcal{Y}}_{2k}^{(\tau)}}{\mu^{(\tau)}})\big)}{2(N-1)}\!\right)\\
		&+P_{\Omega}(\dot{\mathcal{T}}).
	\end{split}
\end{equation}
 
\textbf{Updating $\{\dot{\mathcal{M}}_{k}\}_{k=1}^{N-1}$}: In the $(\tau+1)$-th iteration, fixing the other variables at their latest values, each $\dot{\mathcal{M}}_{k}^{(\tau+1)}$ is related to solving the following optimization problem:
\begin{equation}\label{equupm1}\small
\begin{split}
		\dot{\mathcal{M}}_{k}^{(\tau+1)}=&\mathop{{\rm{arg\, min}}}\limits_{\dot{\mathcal{M}}_{k}}\ \alpha_{k}\|\dot{\mathcal{M}}_{k[k]}\|_{w,\ast}\\
		&+\mathfrak{R}(\langle\dot{\mathcal{Y}}_{1k}^{(\tau)},\dot{\mathcal{X}}^{(\tau+1)}-\dot{\mathcal{M}}_{k}\rangle)\\
		&+\frac{\mu^{(\tau)}}{2}\|\dot{\mathcal{X}}^{(\tau+1)}-\dot{\mathcal{M}}_{k}\|_{F}^{2}\\
		=&\mathop{{\rm{arg\, min}}}\limits_{\dot{\mathcal{M}}_{k}}\ \frac{\alpha_{k}}{\mu^{(\tau)}}\|\dot{\mathcal{M}}_{k[k]}\|_{w,\ast}\\
		&+\frac{1}{2}\|\dot{\mathcal{M}}_{k}-(\dot{\mathcal{X}}^{(\tau+1)}+\frac{\dot{\mathcal{Y}}_{1k}^{(\tau)}}{\mu^{(\tau)}})\|_{F}^{2}.
\end{split}
\end{equation}
Based on the equation that $\|\dot{\mathcal{X}}\|_{F}=\|\dot{\mathcal{X}}_{[k]}\|_{F}$, (\ref{equupm1}) can be rewritten as
\begin{equation}\label{equupm2}\small
	\begin{split}
		\dot{\mathcal{M}}_{k}^{(\tau+1)}
		=&\mathop{{\rm{arg\, min}}}\limits_{\dot{\mathcal{M}}_{k}}\ \frac{\alpha_{k}}{\mu^{(\tau)}}\|\dot{\mathcal{M}}_{k[k]}\|_{w,\ast}\\
		&+\frac{1}{2}\|\dot{\mathcal{M}}_{k[k]}-(\dot{\mathcal{X}}^{(\tau+1)}_{[k]}+\frac{\dot{\mathcal{Y}}_{1k[k]}^{(\tau)}}{\mu^{(\tau)}})\|_{F}^{2}.
	\end{split}
\end{equation}
Denote $\dot{\mathbf{\Gamma}}$:= $\dot{\mathcal{X}}^{(\tau+1)}_{[k]}+\frac{\dot{\mathcal{Y}}_{1k[k]}^{(\tau)}}{\mu^{(\tau)}}$ and let $\dot{\mathbf{\Gamma}}=\dot{\mathbf{U}}_{k}\mathbf{\Sigma}_{k}\dot{\mathbf{V}}_{k}^{H}$ be the QSVD of $\dot{\mathbf{\Gamma}}$, where 
\begin{equation*}\small
\mathbf{\Sigma}_{k}=\left[\begin{array}{cc}
	{\rm{diag}}\big(\sigma_{1}(\dot{\mathbf{\Gamma}}), \ldots, \sigma_{s}(\dot{\mathbf{\Gamma}})\big)\\ 
	 \mathbf{0}
\end{array} \right],
\end{equation*}
and $\sigma_{n}(\dot{\mathbf{\Gamma}})$ is the $n$-th singular value of $\dot{\mathbf{\Gamma}}$, $s$ denotes the number of nonzero singular values of $\dot{\mathbf{\Gamma}}$. From \cite{DBLP:journals/ijon/YuZY19}, the problem (\ref{equupm2}) has the following closed-form solution:
\begin{equation}\label{equupm3}\small
		\dot{\mathcal{M}}_{k}^{(\tau+1)}
		={\rm{fold}}_{[k]}(\dot{\mathbf{U}}_{k}\hat{\mathbf{\Sigma}}_{k}\dot{\mathbf{V}}_{k}^{H}),
\end{equation}
where 
\begin{equation*}\small
	\hat{\mathbf{\Sigma}}_{k}=\left[\begin{array}{cc}
		{\rm{diag}}\big(\sigma_{1}(\dot{\mathcal{M}}_{k[k]}^{(\tau+1)}),\ldots, \sigma_{s}(\dot{\mathcal{M}}_{k[k]}^{(\tau+1)})\big)\\ 
		\mathbf{0}
	\end{array} \right],
\end{equation*}
and 
\begin{equation*}\small
\sigma_{n}(\dot{\mathcal{M}}_{k[k]}^{(\tau+1)})=	\left\{
	\begin{array}{lc}
		0,\qquad &\text{if}\  c_{2}<0\\
		\frac{c_{1}+\sqrt{c_{2}}}{2}, &\text{if}\  c_{2}\geq0
	\end{array},
	\right.
\end{equation*}
where $c_{1}=\sigma_{n}(\dot{\mathbf{\Gamma}})-\epsilon$, $c_{2}=(\sigma_{n}(\dot{\mathbf{\Gamma}})+\epsilon)^{2}-4C$, and $C$ is a compromising constant.

\textbf{Updating $\{\dot{\mathcal{E}}_{k}\}_{k=1}^{N-1}$}: In the $(\tau+1)$-th iteration, fixing the other variables at their latest values, each $\dot{\mathcal{E}}_{k}^{(\tau+1)}$ is related to solving the following optimization problem:
\begin{equation}\label{equupe1}\small
	\begin{split}
		\dot{\mathcal{E}}_{k}^{(\tau+1)}=&\mathop{{\rm{arg\, min}}}\limits_{\dot{\mathcal{E}}_{k}}\ \lambda_{k}\|\dot{\mathcal{E}}_{k}\|_{1}\\
		&+\mathfrak{R}(\langle\dot{\mathcal{Y}}_{2k}^{(\tau)},\dot{\mathcal{E}}_{k}-\mathfrak{T}(\dot{\mathcal{X}}^{(\tau+1)})\rangle)\\
		&+\frac{\mu^{(\tau)}}{2}\|\dot{\mathcal{E}}_{k}-\mathfrak{T}(\dot{\mathcal{X}}^{(\tau+1)})\|_{F}^{2}\\
		=&\mathop{{\rm{arg\, min}}}\limits_{\dot{\mathcal{E}}_{k}}\ \frac{\lambda_{k}}{\mu^{(\tau)}}\|\dot{\mathcal{E}}_{k}\|_{1}\\
		&+\frac{1}{2}\|\dot{\mathcal{E}}_{k}-(\mathfrak{T}(\dot{\mathcal{X}}^{(\tau+1)})-\frac{\dot{\mathcal{Y}}_{2k}^{(\tau)}}{\mu^{(\tau)}})\|_{F}^{2}\\
		=&{\rm{shinkQ}}(\mathfrak{T}(\dot{\mathcal{X}}^{(\tau+1)})-\frac{\dot{\mathcal{Y}}_{2k}^{(\tau)}}{\mu^{(\tau)}},\frac{\lambda_{k}}{\mu^{(\tau)}}),
	\end{split}
\end{equation}
where, for any quaternion matrix $\dot{\mathbf{X}}=[\dot{x}_{mn}]\in \mathbb{H}^{M\times N}$ and $\gamma>0$,  ${\rm{shinkQ}}(\dot{\mathbf{X}},\gamma)$ is defined as \cite{DBLP:journals/tip/JiaJNZ22}
\begin{equation*}\small
{\rm{shinkQ}}(\dot{\mathbf{X}},\gamma)={\rm{signQ}}(\dot{x}_{mn}) \max(|\dot{x}_{mn}|-\gamma,0),	
\end{equation*}
and 
\begin{equation*}\small
	{\rm{signQ}}(\dot{x}_{mn})=	\left\{
	\begin{array}{lc}
		\frac{\dot{x}_{mn}}{|\dot{x}_{mn}|},\qquad &\text{if}\  |\dot{x}_{mn}|\neq0,\\
		0, &\text{otherwise}.
	\end{array}
	\right.
\end{equation*}

\textbf{Updating $\{\dot{\mathcal{Y}}_{1k}\}_{k=1}^{N-1}$, $\{\dot{\mathcal{Y}}_{2k}\}_{k=1}^{N-1}$, and $\mu$}: In the $(\tau+1)$-th iteration, fixing the other variables at their latest values, each $\dot{\mathcal{Y}}_{1k}^{(\tau+1)}$, each $\dot{\mathcal{Y}}_{2k}^{(\tau+1)}$, and $\mu^{(\tau+1)}$ are respectively updated by:
\begin{equation}\label{equy1}\small
\dot{\mathcal{Y}}_{1k}^{(\tau+1)}=\dot{\mathcal{Y}}_{1k}^{(\tau)}+\mu^{(\tau)}(\dot{\mathcal{X}}^{(\tau+1)}-\dot{\mathcal{M}}_{k}^{(\tau+1)}),
\end{equation}
\begin{equation}\label{equy2}\small
\dot{\mathcal{Y}}_{2k}^{(\tau+1)}=\dot{\mathcal{Y}}_{2k}^{(\tau)}+\mu^{(\tau)}(\mathfrak{T}(\dot{\mathcal{X}}^{(\tau+1)})-\dot{\mathcal{E}}_{k}^{(\tau+1)}),	
\end{equation}
\begin{equation}\label{equmu}\small
\mu^{(\tau+1)}=	\min(\rho\mu^{(\tau)}, \mu_{\max}),
\end{equation}
where $\rho>1$  is a constant parameter, and $\mu_{\max}$ is the default maximum of $\mu$.

Finally, the algorithm of the \textbf{Q}uaternion \textbf{T}ensor \textbf{T}rain rank minimization with \textbf{S}parse \textbf{R}egularization in a \textbf{T}ransformed \textbf{D}omain for quaternion tensor completion (QTT-SRTD) is summarized in Table \ref{tab_algorithm}. 
\begin{table}[htbp]
	\caption{The calculation procedure of the proposed QTT-SRTD.}
	\hrule
	\label{tab_algorithm}
	\begin{algorithmic}[1]
		\REQUIRE The observed $N$-th order quaternion tensor $\dot{\mathcal{T}}\in \mathbb{H}^{I_{1}\times I_{2}\times \ldots \times I_{N}}$ with $\Omega$ (the index of observed entries), unitary quaternion
		transform matrices $\{\dot{\mathbf{T}}_{n}\}_{n=1}^{N}$, $\{\lambda_{k}\}_{k=1}^{N-1}$, $\{\alpha_{k}\}_{k=1}^{N-1}$, $\mu_{\max}$ and $\rho$.
		\STATE \textbf{Initialize} $\{\dot{\mathcal{M}}_{k}^{(0)}\}_{k=1}^{N-1}$, $\{\dot{\mathcal{E}}_{k}^{(0)}\}_{k=1}^{N-1}$, $\{\dot{\mathcal{Y}}_{1k}^{(0)}\}_{k=1}^{N-1}$, $\{\dot{\mathcal{Y}}_{2k}^{(0)}\}_{k=1}^{N-1}$, $\mu^{(0)}$, and $\tau=0$.
		\STATE \textbf{Repeat}
		\STATE Update $\dot{\mathcal{X}}^{(\tau+1)}$ via (\ref{equupx}). 
		\FOR {$k=1$ to $N-1$}
		\STATE Update $\dot{\mathcal{M}}_{k}^{(\tau+1)}$ via (\ref{equupm3});
		\STATE Update $\dot{\mathcal{E}}_{k}^{(\tau+1)}$ via (\ref{equupe1});
		\STATE Update $\dot{\mathcal{Y}}_{1k}^{(\tau+1)}$ via (\ref{equy1});
		\STATE Update $\dot{\mathcal{Y}}_{2k}^{(\tau+1)}$ via (\ref{equy2}).
		\ENDFOR
		\STATE Update $\mu^{(\tau+1)}$ via (\ref{equmu}).
		\STATE  $\tau\longleftarrow \tau+1$.
		\STATE \textbf{Until} $\frac{\|\dot{\mathcal{X}}^{(\tau+1)}-\dot{\mathcal{X}}^{(\tau)}\|_{F}}{\|\dot{\mathcal{T}}\|_{F}}<10^{-5}$.
		\ENSURE  \text{The recovered quaternion tensor}\  $\dot{\mathcal{X}}^{(\tau)}$.
	\end{algorithmic}
	\hrule
\end{table}

\section{Quaternion Tensor Augmentation}\label{sec_QKA}
The ket augmentation (KA) \cite{DBLP:journals/tip/BenguaPTD17} as a tensor augmentation technique, has been proved it explores the low-rank structure more obviously than the original one. Thus, KA is a helpful pretreatment step for TT rank-based optimization \cite{yang2020tensor,DBLP:conf/icspcs/BenguaTPD16}. 
 
In \cite{DBLP:journals/tip/BenguaPTD17}, the authors treat color images as third-order tensors whose third dimension is $3$ and defined KA. In contrast, we treat color images as second-order quaternion tensors (quaternion matrices) and define quaternion KA (QKA) to
represent a lower-order quaternion tensor by a higher-order one. For example, QKA reshapes
a color image $\dot{\mathbf{G}}\in\mathbb{H}^{M_{1}\times M_{2}}$ ($M_{1}\times M_{2}=2^{n}\times 2^{n}$ is the number of pixels in the image) into an $N$-th order quaternion tensor $\dot{\mathcal{Q}}\in \mathbb{H}^{I_{1}\times I_{2}\times \ldots \times I_{N}}$, where $M_{1}M_{2}=\Pi_{j=1}^{N}I_{j}$,  and $I_{j}$ represents a unique block structured addressing of the original color image. In the following, we give the details of the structured block addressing procedure of QKA.

For a color image $\dot{\mathbf{G}}\in\mathbb{H}^{M_{1}\times M_{2}}$, we consider the initial
smallest block (labeled as $i_{1}$) with size $2\times 2$, which can be represented as
\begin{equation}\label{qka1}\small
\dot{\mathbf{G}}_{[2^{1}\times 2^{1}]}=\sum_{i_{1}=1}^{4}\dot{q}_{i_{1}}\mathbf{e}_{i_{1}},	
\end{equation}
where $\dot{q}_{i_{1}}$ is the color pixel value, $\mathbf{e}_{i_{1}}$ is the orthonormal base, \emph{i.e.}, $\mathbf{e}_{1}=(1,0,0,0)$, $\mathbf{e}_{2}=(0,1,0,0)$, $\mathbf{e}_{3}=(0,0,1,0)$, and $\mathbf{e}_{4}=(0,0,0,1)$. The value $i_{1}=1$ can be understood as labeling the up-left color pixel, $i_{1}=2$ as the up-right one, $i_{1}=3$ as down-left one, and $i_{1}=4$ as down-right one, \emph{see} Figure \ref{qka_fig} (left-hand side ). We then consider a larger block (labeled as $i_{2}$) with size $4\times 4$ made up of four inner sub-blocks as shown Figure \ref{qka_fig} (right-hand side). The new block is represented by
\begin{equation}\label{qka2}\small
	\dot{\mathbf{G}}_{[2^{2}\times 2^{2}]}=\sum_{i_{2}=1}^{4}\sum_{i_{1}=1}^{4}\dot{q}_{i_{2}i_{1}}\mathbf{e}_{i_{2}}\otimes \mathbf{e}_{i_{1}},	
\end{equation}
where $\otimes$ denotes the Kronecker product. Generally, this block structure can be extended to a size of $2^{N} \times 2^{N}$ step by step until it can present all the values of color pixels in the image. Finally, the color image can be cast into an $N$-th order quaternion tensor $\dot{\mathcal{Q}}\in \mathbb{H}^{4\times 4\times \ldots \times 4}$ containing all the color pixel values as follows:
\begin{equation}\label{equqka}\small
	\dot{\mathbf{G}}_{[2^{N}\times 2^{N}]}=\!\!\!\sum_{i_{N},\ldots,i_{2}, i_{1}=1}^{4}\!\!\!\dot{q}_{i_{N}\ldots i_{2}i_{1}}\mathbf{e}_{i_{N}}\otimes\ldots \otimes \mathbf{e}_{i_{2}}\otimes \mathbf{e}_{i_{1}}.	
\end{equation}
\begin{remark}
The presentation (\ref{equqka}) is suitable for color image processing as it preserves all the color pixel values and rearranges them in a higher-order quaternion tensor such that not only the QTT-rank minimization method can be well used to process color images, but also the richness of textures in the
color image can be studied via the correlation between modes of the quaternion
tensor. Figure \ref{example1}(a) shows the process of converting a three-channel color image to an eighth-order quaternion tensor.
\begin{figure}[htbp]
	\centering
	\includegraphics[width=7cm,height=4cm]{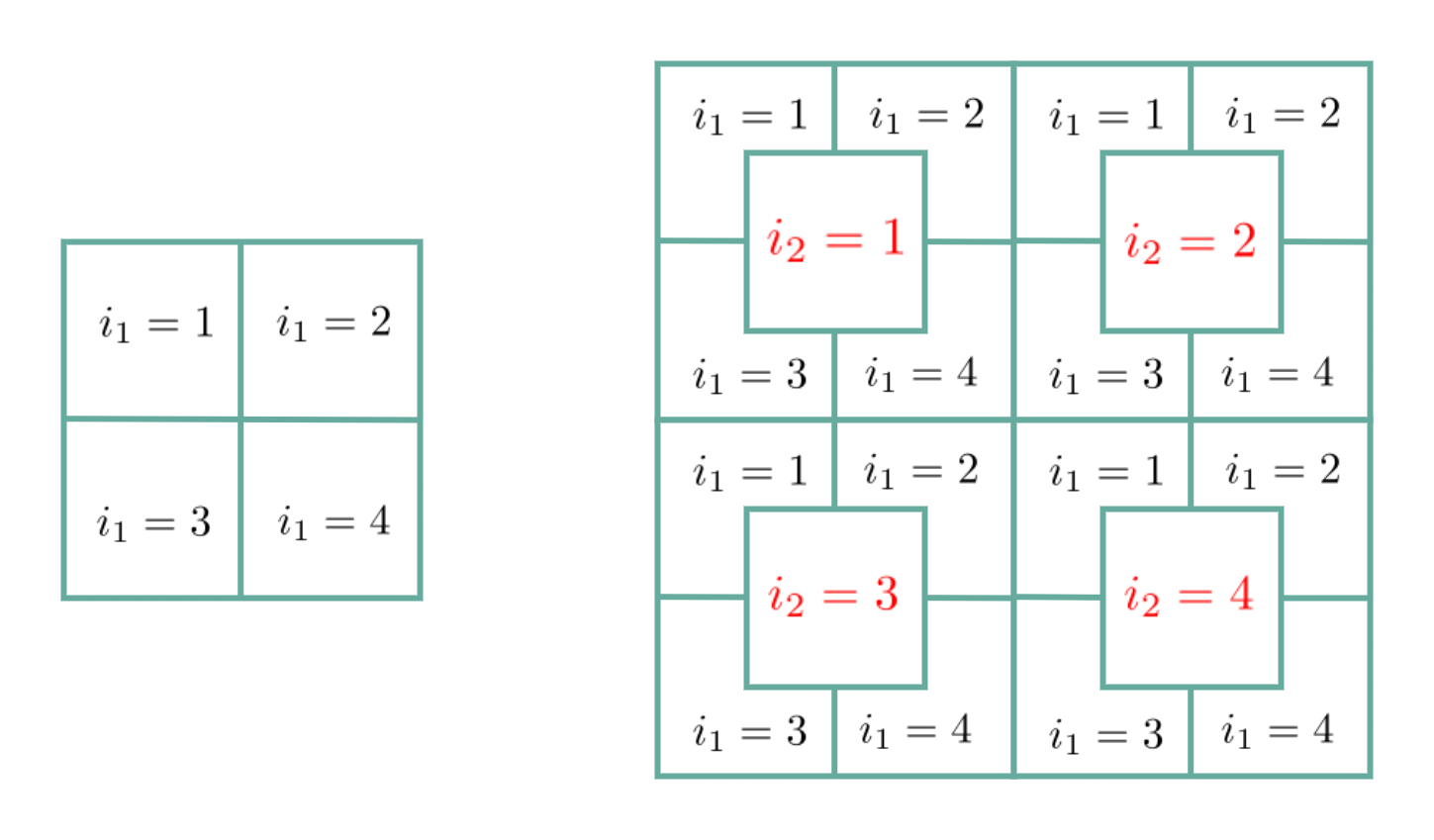}
	\caption{The illustration of QKA casting a color image represented by a quaternion matrix to a higher-order quaternion tensor. The left-hand side is an example of a block with size $2\times 2$ represented by (\ref{qka1}). The right-hand side is an example of a block with size $2^{2}\times 2^{2}$ represented by (\ref{qka2}).}
	\label{qka_fig}
\end{figure}
\end{remark}
\begin{remark}
The defined QKA can also be used to rearrange a color video represented by a third-order quaternion tensor $\dot{\mathcal{V}}\in\mathbb{H}^{C_{1}\times C_{2}\times F}$ (image row $\times$ image column $\times$ frame) in a higher-order quaternion tensor $\dot{\mathcal{T}}\in\mathbb{H}^{I_{1}\times I_{2}\times \ldots \times I_{N}\times F}$, where $C_{1}C_{2}=\Pi_{j=1}^{N}I_{j}$. The general form is given by
\begin{equation}\label{equqka_v}\small
\begin{split}
	&\dot{\mathcal{V}}_{[2^{N}\times 2^{N}\times F ]}\\
	&=\sum_{i_{N},\ldots, i_{1}=1}^{4}\sum_{f=1}^{F}\dot{t}_{i_{N}\ldots i_{1}}\mathbf{e}_{i_{N}}\otimes\ldots\otimes \mathbf{e}_{i_{1}}\otimes \mathbf{u}_{f},
\end{split}	
\end{equation}
where $\mathbf{u}_{f}$ is also an orthonormal base defined as $\mathbf{u}_{1}=(1,0,\ldots,0), \mathbf{u}_{2}=(0,1,\ldots,0),\ldots,\mathbf{u}_{F}=(0,0,\ldots,1)$.
In addition, it should be noted that in (\ref{equqka}) and (\ref{equqka_v}), $i_{n}=1,\ldots,c$, where $c>1$ can be any appropriate value, i.e., $c=4$ is not necessary. 
\end{remark}

\section{Numerical Experiments}\label{sec_6}
In this section, extensive experiments are conducted to demonstrate the effectiveness and superiority of the proposed QTT-SRTD for inpainting problems of color images and color videos. All the experiments are run in MATLAB $2014b$ under Windows $10$ on a personal computer with a $1.60$GHz CPU and $8$GB memory.

\subsection{Color Image Inpainting}
In this experiment, we apply the proposed QTT-SRTD to color image inpainting from incomplete entries.

\textbf{Compared methods:} We compare the proposed QTT-SRTD with several well-known methods including t-SVD \cite{DBLP:journals/tsp/ZhangA17}, SiLRTC-TT \cite{DBLP:journals/tip/BenguaPTD17}, TMac-TT \cite{DBLP:journals/tip/BenguaPTD17}, LRQA-2 \cite{DBLP:journals/tip/ChenXZ20}, LRQMC \cite{DBLP:journals/tip/MiaoK22}, and TQLNA \cite{DBLP:journals/isci/YangMK22}.

\textbf{Parameter and initialization setting\footnote{All parameter choices and initialization settings are based on empirical and experimental results.}:} For our proposed QTT-SRTD in Table \ref{tab_algorithm}, we choice all quaternion
transform matrices $\{\dot{\mathbf{T}}_{n}\}_{n=1}^{N}$ as QDWHT\footnote{One can try other transforms, but here we just use QDWHT without explanation based on the results of the experiment. The difference between different transforms in specific applications is not the focus of this paper.}, all $\{\lambda_{k}\}_{k=1}^{N-1}$ are seted as $0.01$, all $\{\alpha_{k}\}_{k=1}^{N-1}$ are seted as $\frac{1}{N-1}$, 
$\mu_{\max}=10^{6}$, and $\rho=1.08$. $\{\dot{\mathcal{M}}_{k}^{(0)}\}_{k=1}^{N-1}$, $\{\dot{\mathcal{E}}_{k}^{(0)}\}_{k=1}^{N-1}$, $\{\dot{\mathcal{Y}}_{1k}^{(0)}\}_{k=1}^{N-1}$, and $\{\dot{\mathcal{Y}}_{2k}^{(0)}\}_{k=1}^{N-1}$, are all simply initialized to random quaternion matrices. $\mu^{(0)}$ is initialized to $2.5*10^{-3}$.  In addition, all compared methods are
from the source codes and the parameter settings are based
on the suggestions in the orginal papers.

\begin{figure*}[htbp]
	\centering
	\subfigure[baboon]{
		\includegraphics[width=3.2cm,height=2cm]{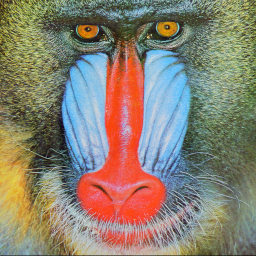}
	}
	\subfigure[house]{
		\includegraphics[width=3.2cm,height=2cm]{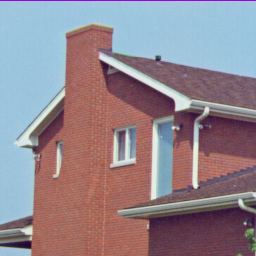}
	}
	\subfigure[airplane]{
		\includegraphics[width=3.2cm,height=2cm]{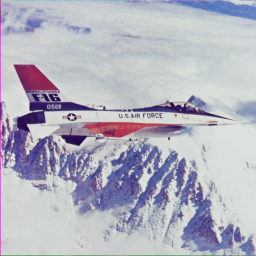}
	}
	\subfigure[peppers]{
	    \includegraphics[width=3.2cm,height=2cm]{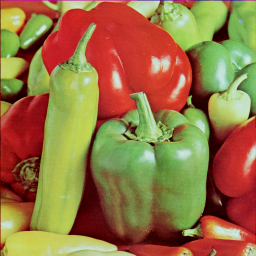}
	}\\
	\subfigure[sailboat]{
		\includegraphics[width=3.2cm,height=2cm]{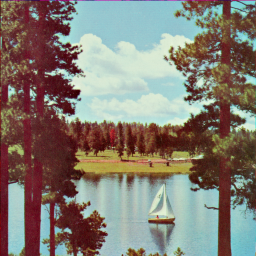}
	}
	\subfigure[lena]{
		\includegraphics[width=3.2cm,height=2cm]{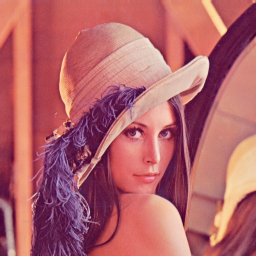}
	}
	\subfigure[panda]{
		\includegraphics[width=3.2cm,height=2cm]{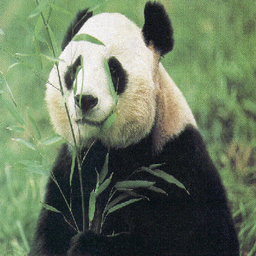}
	}
	\subfigure[barbara]{
		\includegraphics[width=3.2cm,height=2cm]{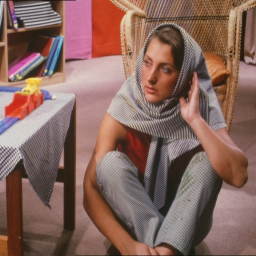}
	}
	\caption{Original color images.}
	\label{Test_data}
\end{figure*}

\begin{figure*}[htbp]
	\centering
	\subfigure[]{
		\includegraphics[width=2cm,height=13cm]{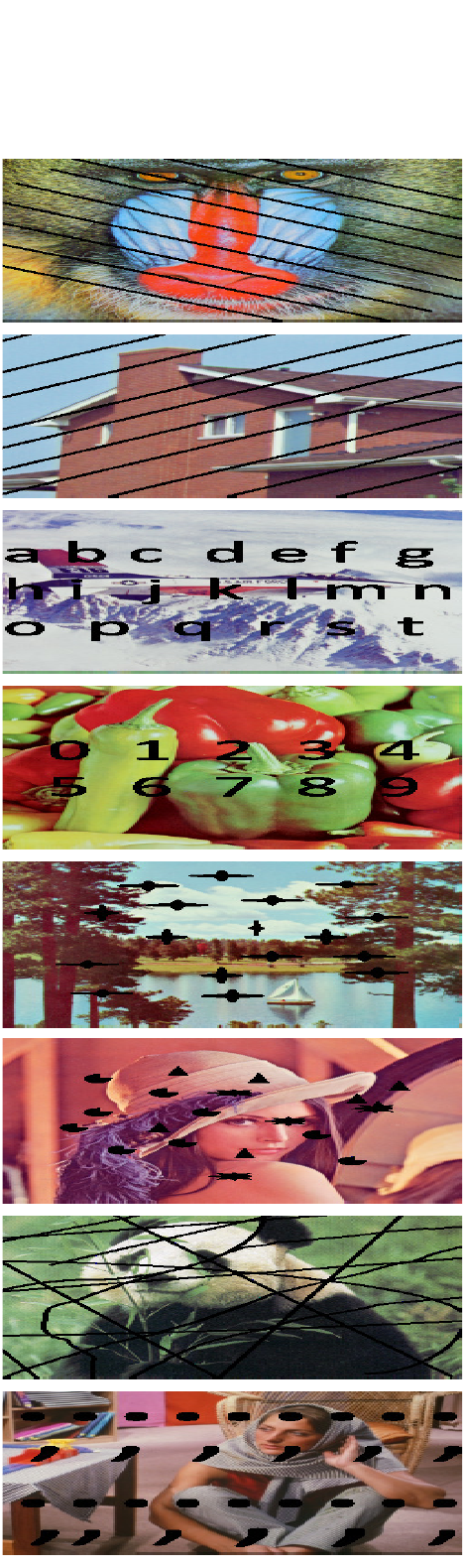}
	}
	\hspace{-0.18in}
	\subfigure[]{
		\includegraphics[width=2cm,height=13cm]{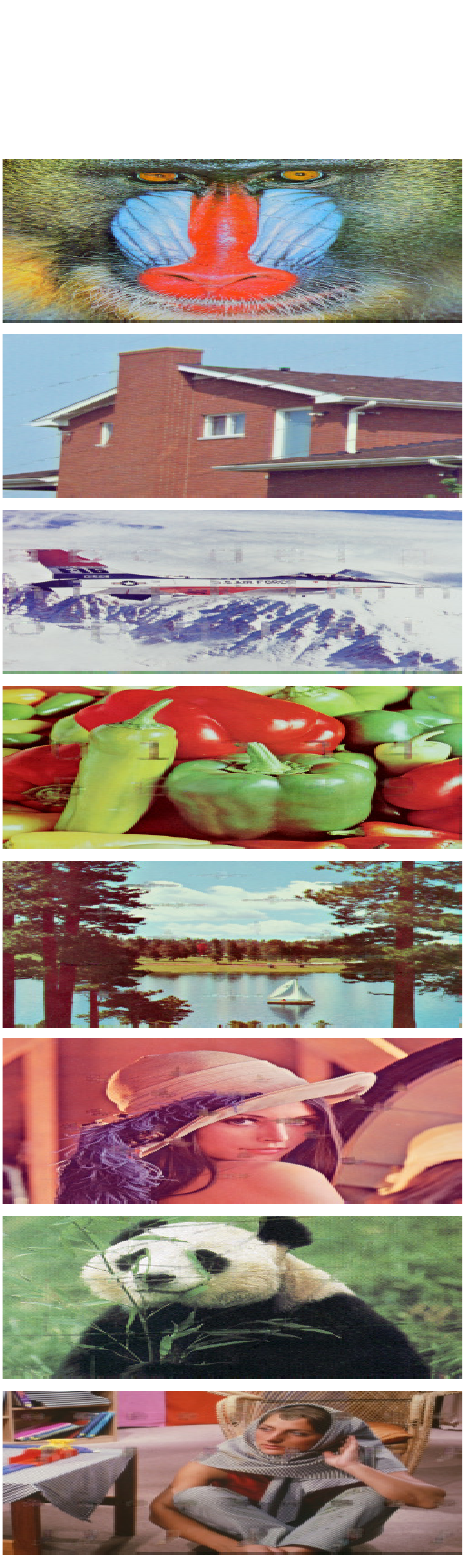}
	}
	\hspace{-0.18in}
	\subfigure[]{
		\includegraphics[width=2cm,height=13cm]{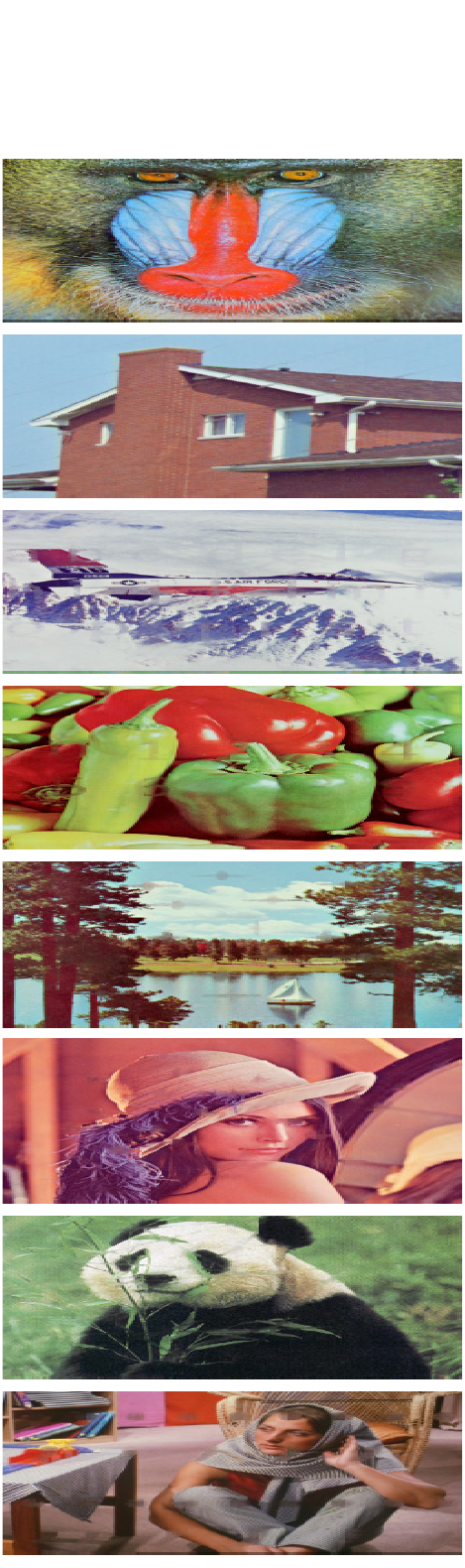}
	}
	\hspace{-0.18in}
	\subfigure[]{
		\includegraphics[width=2cm,height=13cm]{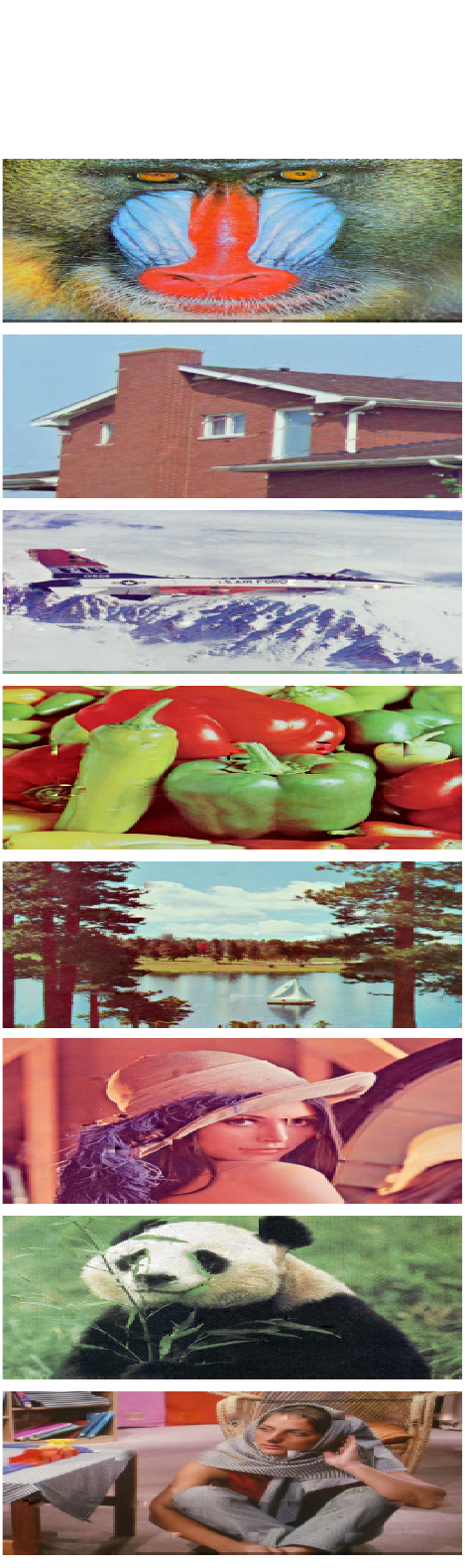}
	}
	\hspace{-0.18in}
	\subfigure[]{
		\includegraphics[width=2cm,height=13cm]{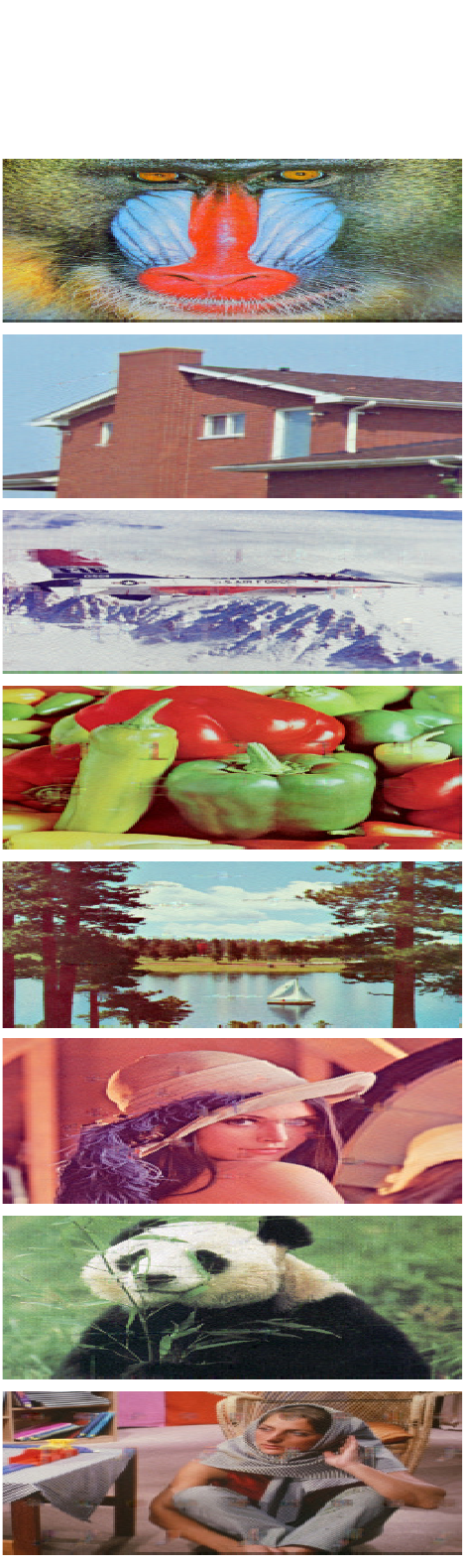}
	}
	\hspace{-0.18in}
	\subfigure[]{
		\includegraphics[width=2cm,height=13cm]{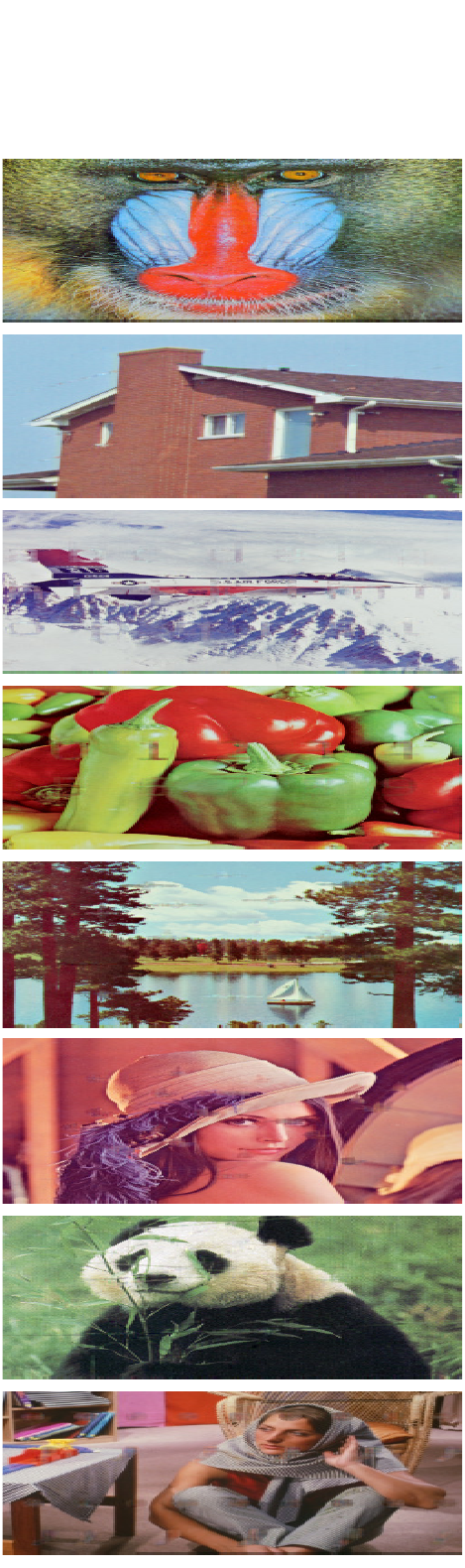}
	}
	\hspace{-0.18in}
	\subfigure[]{
		\includegraphics[width=2cm,height=13cm]{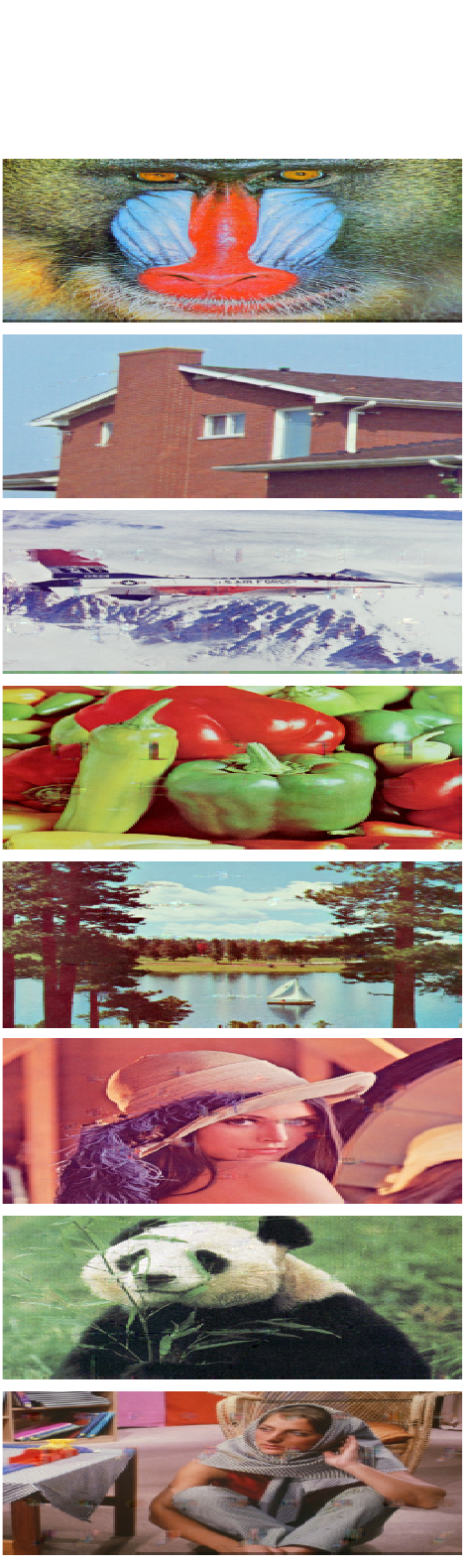}
	}
	\hspace{-0.18in}
	\subfigure[]{
		\includegraphics[width=2cm,height=13cm]{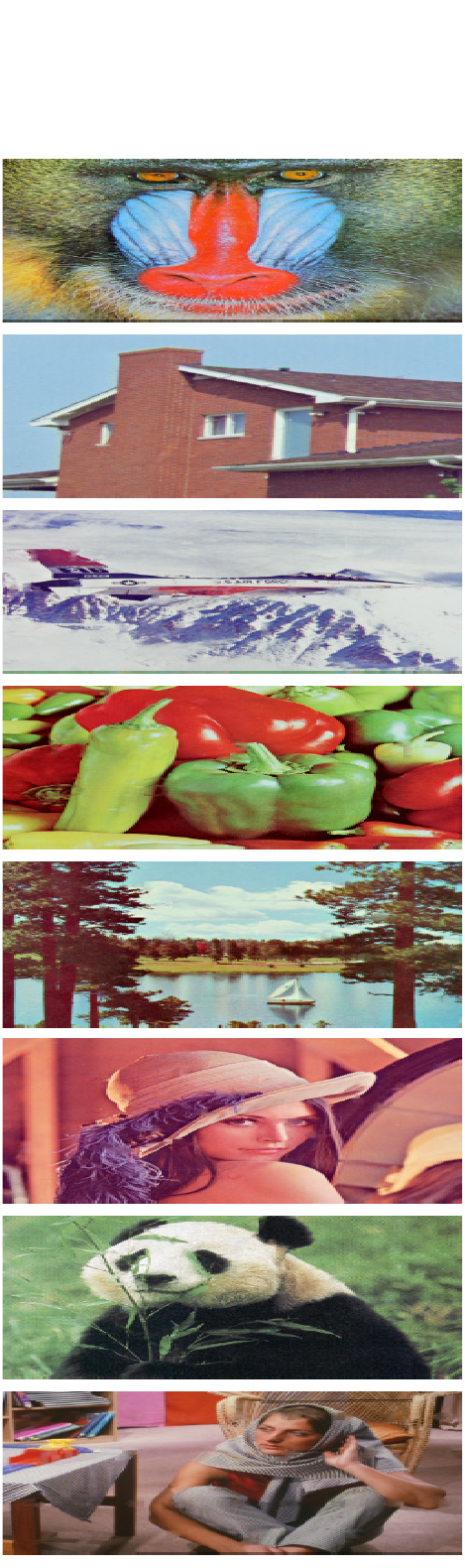}
	}
	\caption{Recovered color images for eight
		kinds of structural missing pixels. From left to right: the observed color images, the recovered results
		by t-SVD, SiLRTC-TT, TMac-TT, LRQA-2, LRQMC, TQLNA, and QTT-SRTD, respectively. From top to bottom: baboon, house, airplane, peppers, sailboat, lena, panda, and barbara. \textbf{The figure is viewed better in zoomed PDF}.
	}
	\label{im_s2}
\end{figure*}

\textbf{Test data and settings:} Numerical comparisons are implemented on eight common-used color images (including ``baboon'', ``house'', ``airplane'', ``peppers'', ``sailboat'', ``lena'', ``panda'', and ``barbara'', \emph{see} Figure \ref{Test_data}) with size $256\times 256$. For the
 QTT-SRTD, the color images are transformed into eighth-order quaternion tensors with size $4\times 4\times 4\times 4\times 4\times 4\times 4\times 4$ by QKA. In addition, for random missing, we set four levels of sampling rates (SRs) which are ${\rm{SR}}=10\%$, ${\rm{SR}}=20\%$, ${\rm{SR}}=30\%$, and ${\rm{SR}}=40\%$.
 For structural missing, we use eight kinds of structural missing pixels, \emph{see} the first column in Figure \ref{im_s2}.
\begin{table*}[htbp]
	\caption{The PSNR and SSIM values of different methods on the eight color images with four levels of sampling rates (the format is PSNR/SSIM, and \textbf{bold} fonts denote the best performance).}
	\centering
	\resizebox{16.4cm}{8cm}{
		\begin{tabular}{|c|c|c|c|c|c|c|c|}		
			\hline
			Methods:&t-SVD \cite{DBLP:journals/tsp/ZhangA17} &SiLRTC-TT \cite{DBLP:journals/tip/BenguaPTD17}& TMac-TT \cite{DBLP:journals/tip/BenguaPTD17}& LRQA-2 \cite{DBLP:journals/tip/ChenXZ20}& LRQMC \cite{DBLP:journals/tip/MiaoK22} &TQLNA \cite{DBLP:journals/isci/YangMK22} &\textbf{QTT-SRTD} \\ \toprule
			\hline
			Images:  &\multicolumn{7}{c|}{${\rm{SR}}=10\%$}\\
			\hline
			baboon &17.325/0.505&18.484/0.563&19.242/0.579&17.941/0.527&18.065/0.546&18.075/0.537&\textbf{20.445}/\textbf{0.656}\\
			 house&19.247/0.696&21.239/0.789&22.582/0.817&20.069/0.718&20.091/0.730&19.908/0.716 &\textbf{24.660}/\textbf{0.883}  \\
			airplane&17.765/0.377&20.025/0.596&20.503/0.588&18.601/0.424&18.715/0.480&18.442/0.416   &\textbf{22.337}/\textbf{0.753}  \\
			peppers&15.993/0.718&18.894/0.842&20.416/0.871&17.111/0.761&16.206/0.740&17.145/0.764  &\textbf{22.815}/\textbf{0.920} \\  
			sailboat&16.514/0.502&17.920/0.618&17.606/0.670&17.103/0.550&17.295/0.581&16.711/0.526&\textbf{20.355}/\textbf{0.783}  \\
			 lena&17.659/0.798&20.804/0.888&21.616/0.877&18.607/0.811&18.553/0.825&18.528/0.802&\textbf{23.982}/\textbf{0.932} \\
			panda&18.074/0.492&21.188/0.657&22.518/0.628&19.205/0.514&19.212/0.562&19.149/0.513     &\textbf{24.425}/\textbf{0.770}\\
			 barbara&16.894/0.513&19.834/0.691&20.373/0.723&17.917/0.533&17.947/0.585&17.996/0.527  &\textbf{22.840}/\textbf{0.786}\\
			\hline
			\hline
			Images:  &\multicolumn{7}{c|}{${\rm{SR}}=20\%$}\\
			\hline
			baboon&19.303/0.622&20.423/0.677&20.915/0.696&19.530/0.623&19.726/0.646&19.830/0.640&\textbf{21.690}/\textbf{0.729}\\
			house&22.492/0.827&23.954/0.871&23.788/0.856&23.193/0.836&23.320/0.849&23.335/0.843&\textbf{27.507}/\textbf{0.932}\\
			airplane &20.441/0.555&22.185/0.739&22.024/0.721&21.014/0.580&21.160/0.629&21.114/0.592&\textbf{24.632}/\textbf{0.839} \\
			peppers&19.424/0.846&21.981/0.905&23.059/0.916&20.732/0.874&20.699/0.879&21.086/0.883&\textbf{25.791}/\textbf{0.956} \\  
			sailboat&18.858/0.663&20.218/0.762&21.100/0.794&19.472/0.693&19.570/0.716&19.567/0.699&\textbf{22.602}/\textbf{0.858}\\
			lena&20.859/0.876&23.594/0.927&23.442/0.914&21.698/0.886&21.901/0.897&21.804/0.889&\textbf{26.200}/\textbf{0.955}\\
			 panda&21.397/0.639&23.901/0.761&23.774/0.691&22.307/0.657&22.373/0.690&22.505/0.669&\textbf{26.220}/\textbf{0.825}\\
			 barbara&20.199/0.671&22.619/0.791&23.439/0.807&21.155/0.691&20.739/0.700&21.381/0.698&\textbf{25.244}/\textbf{0.855} \\
			\hline	
			\hline
			Images:  &\multicolumn{7}{c|}{${\rm{SR}}=30\%$}\\
			\hline
			baboon&20.657/0.703 &21.656/0.749&21.801/0.751&20.685/0.695&21.279/0.727&20.878/0.705&\textbf{22.684}/\textbf{0.784}\\
			house&24.944/0.887 &26.106/0.913&26.849/0.908&25.428/0.889&25.640/0.900&25.836/0.899&\textbf{29.592}/\textbf{0.954}\\
			airplane &22.555/0.671 &23.943/0.820&24.096/0.821&22.982/0.681&23.183/0.724&23.250/0.702&\textbf{26.416}/\textbf{0.887}\\
			peppers&22.287/0.908 &24.293/0.939&25.458/0.951&23.330/0.923&23.671/0.930&23.976/0.934&\textbf{27.941}/\textbf{0.973}\\  
			sailboat&20.958/0.767 &22.111/0.838&22.819/0.862&21.343/0.779&21.634/0.806&21.609/0.792&\textbf{24.303}/\textbf{0.900}\\
			lena&23.217/0.917&25.595/0.949&26.136/0.951&23.729/0.921&24.173/0.931&24.059/0.927&\textbf{28.022}/\textbf{0.968} \\
			 panda&23.698/0.730 &25.684/0.821&26.602/0.837&24.297/0.739&24.479/0.772&24.721/0.753&\textbf{27.715}/\textbf{0.863}\\
			 barbara&22.737/0.771 &24.560/0.849&25.338/0.866&23.403/0.781&23.584/0.799&23.885/0.797&\textbf{26.902}/\textbf{0.891}\\
			\hline
						\hline
			Images:  &\multicolumn{7}{c|}{${\rm{SR}}=40\%$}\\
			\hline
			baboon&21.837/0.764 &22.814/0.804&22.621/0.796&21.787/0.756&22.414/0.785&21.936/0.763&\textbf{23.696}/\textbf{0.830}\\
			house&27.360/0.929 &27.995/0.940&28.944/0.941&27.744/0.929&28.140/0.939&28.403/0.938&\textbf{31.329}/\textbf{0.968}\\
			airplane &24.501/0.761 &25.634/0.871&26.184/0.878&24.754/0.757&24.992/0.798&25.208/0.781&\textbf{28.071}/\textbf{0.917}\\
			peppers&24.772/0.943 &26.208/0.959&27.160/0.967&25.624/0.952&26.153/0.958&26.299/0.959&\textbf{29.568}/\textbf{0.981}\\  
			sailboat&22.782/0.837 &23.795/0.888&24.176/0.900&23.099/0.844&23.321/0.861&23.472/0.855&\textbf{25.850}/\textbf{0.928}\\
			lena&25.221/0.943&27.291/0.964&27.726/0.965&25.624/0.945&26.126/0.953&26.061/0.950&\textbf{29.563}/\textbf{0.977}\\
			 panda&25.653/0.799 &27.187/0.863&28.168/0.878&25.985/0.799&26.387/0.829&26.484/0.814&\textbf{28.987}/\textbf{0.892}\\
			 barbara&24.820/0.838 &26.213/0.889&26.690/0.897&25.321/0.843&25.749/0.862&25.825/0.856&\textbf{28.410}/\textbf{0.918}\\
			\hline
	\end{tabular}}
	\label{table_1}
\end{table*}
\begin{figure*}[htbp]
	\centering
	\subfigure[Average PSNR]{\includegraphics[width=8.6cm,height=6cm]{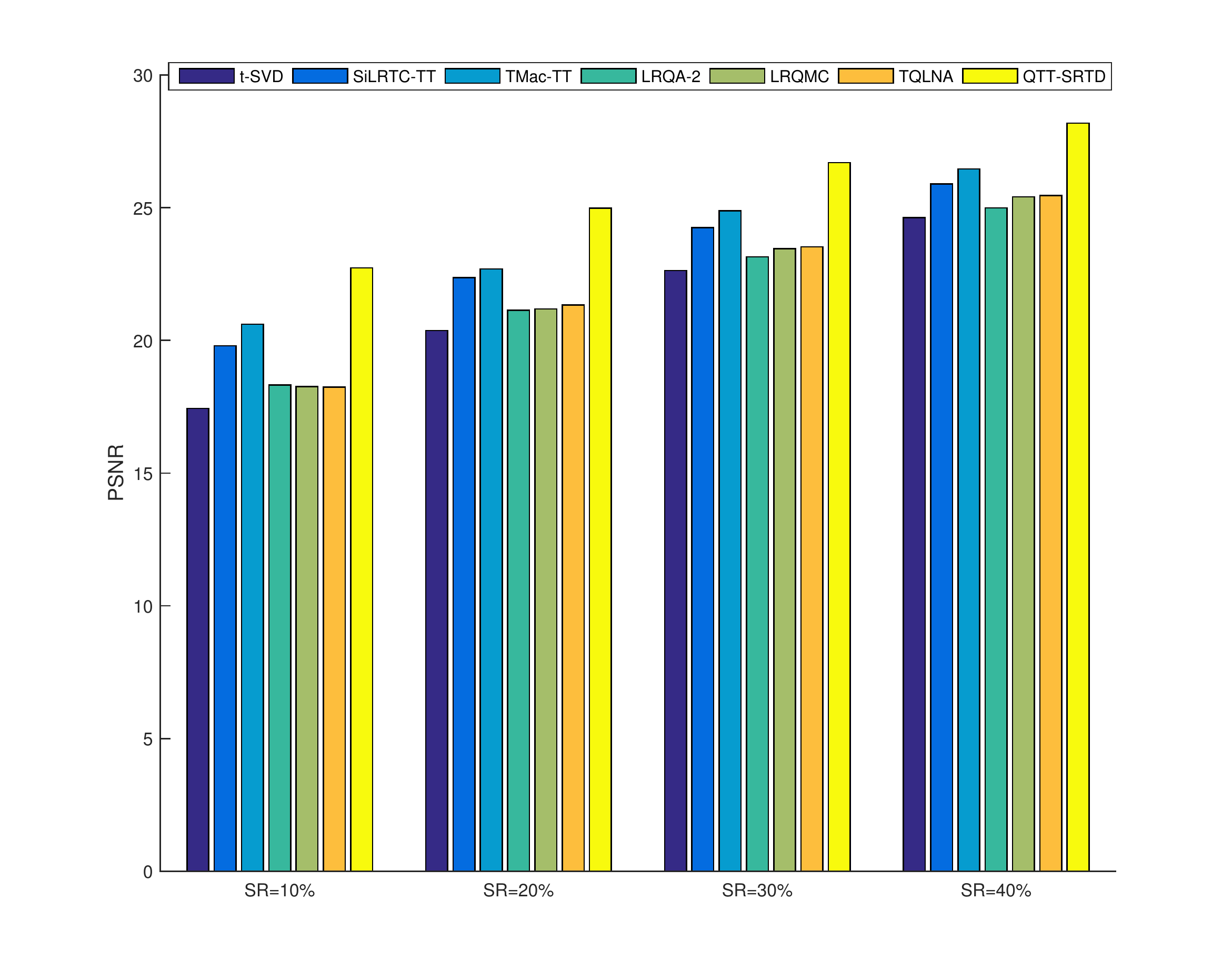}}\hspace{-0.9cm}
	\subfigure[Average SSIM]{\includegraphics[width=8.6cm,height=6cm]{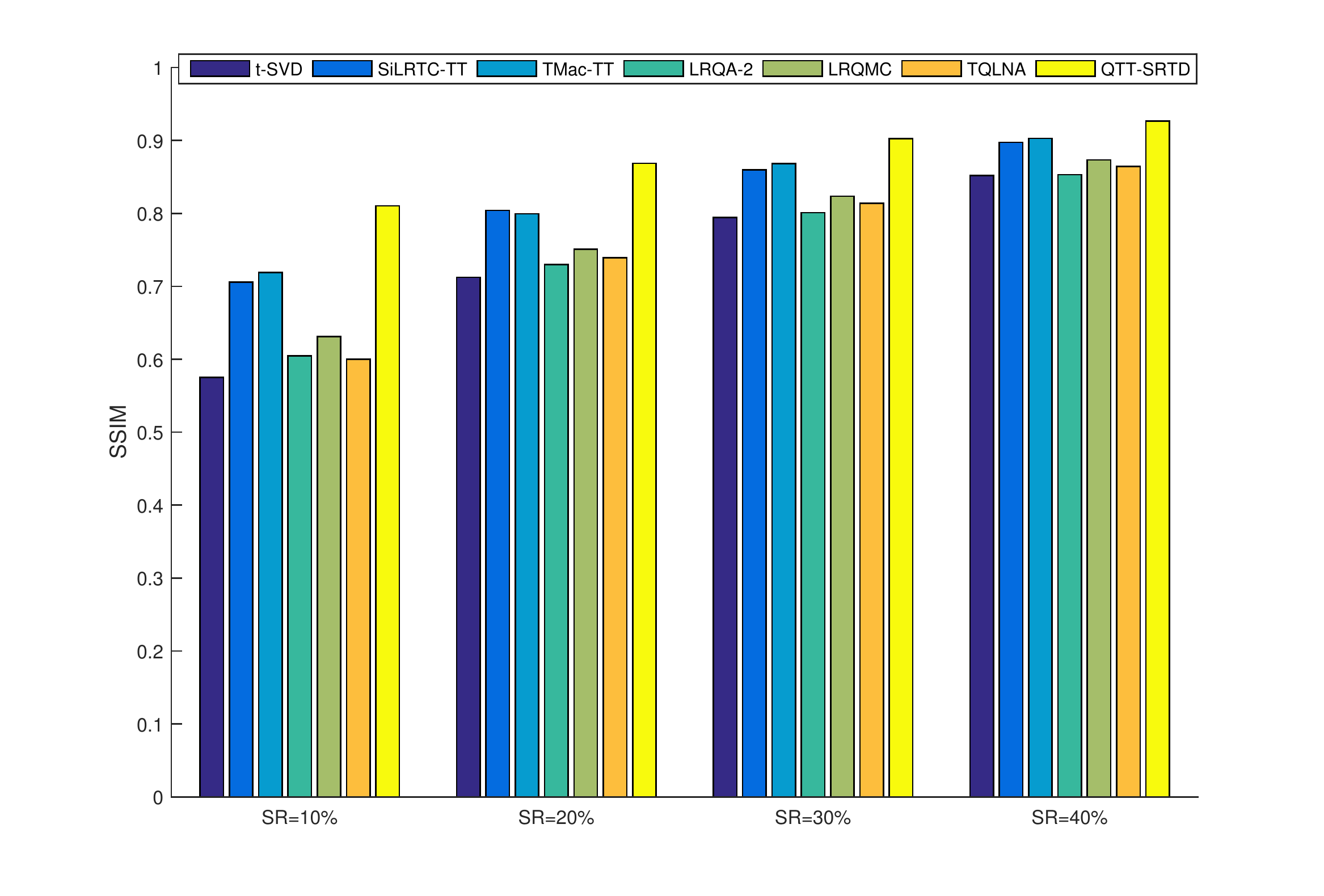}}
	\caption{Average PSNR and SSIM values of different methods on the eight color images with four levels of sampling rates.}
	\label{aps_ssi_1}
\end{figure*}
\begin{figure*}[htbp]
	\centering
	\subfigure[]{
		\includegraphics[width=2cm,height=13cm]{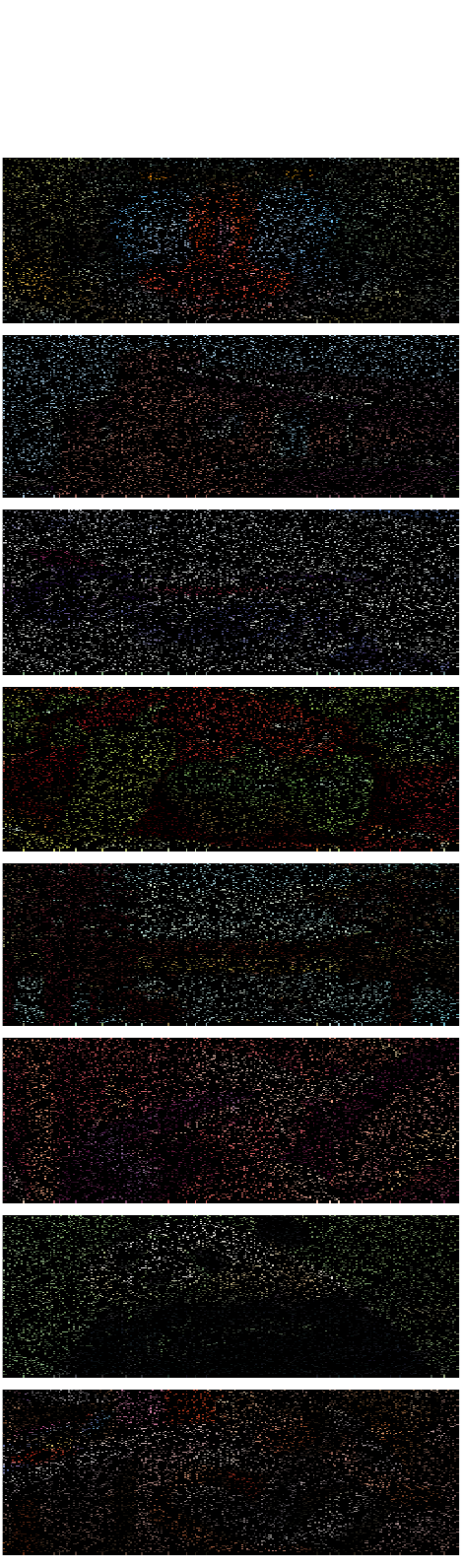}
	}
	\hspace{-0.18in}
	\subfigure[]{
		\includegraphics[width=2cm,height=13cm]{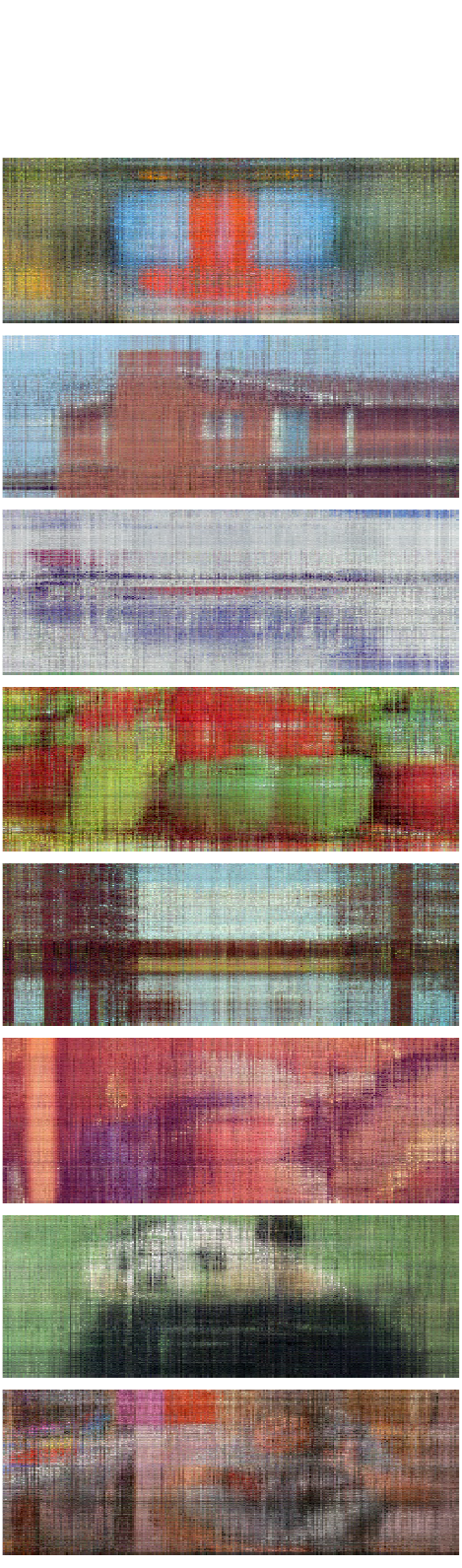}
	}
	\hspace{-0.18in}
	\subfigure[]{
		\includegraphics[width=2cm,height=13cm]{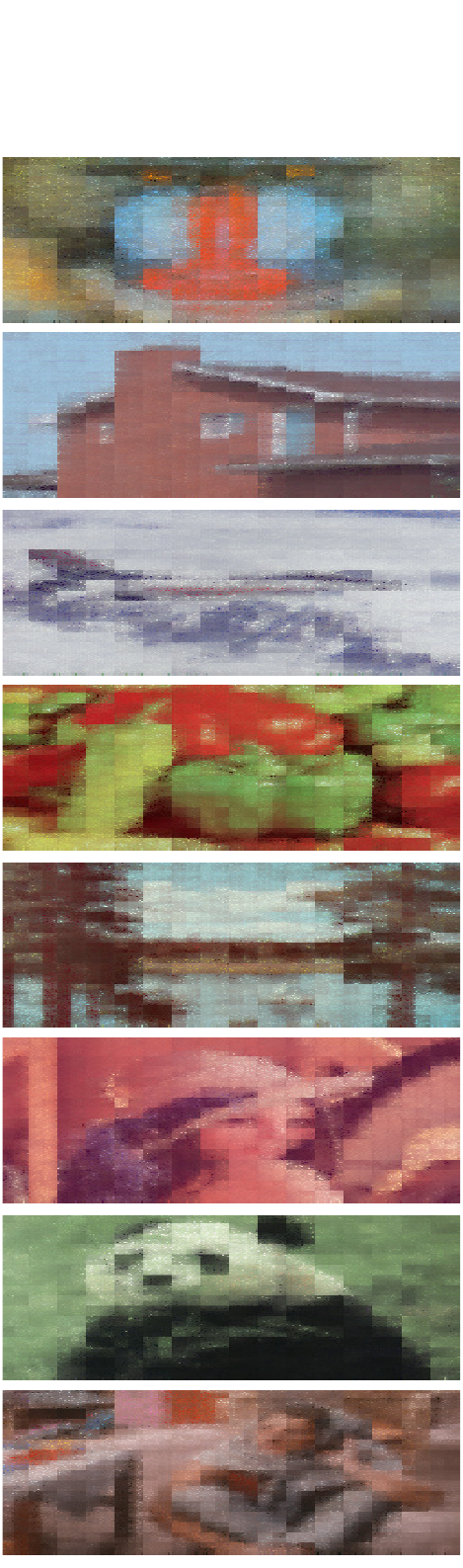}
	}
	\hspace{-0.18in}
	\subfigure[]{
		\includegraphics[width=2cm,height=13cm]{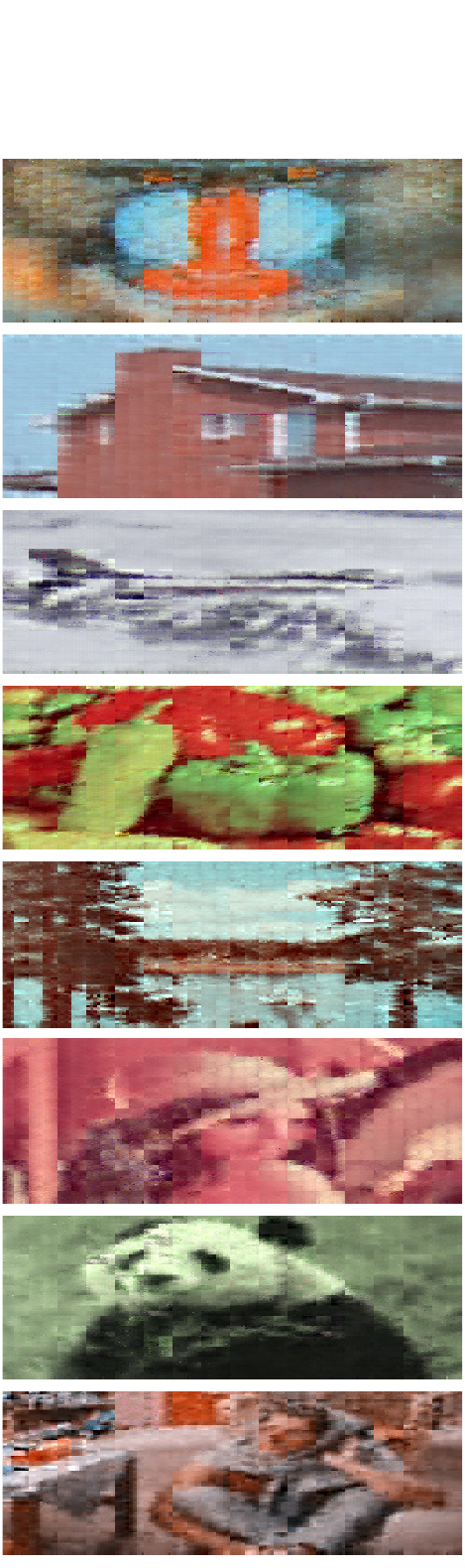}
	}
	\hspace{-0.18in}
	\subfigure[]{
		\includegraphics[width=2cm,height=13cm]{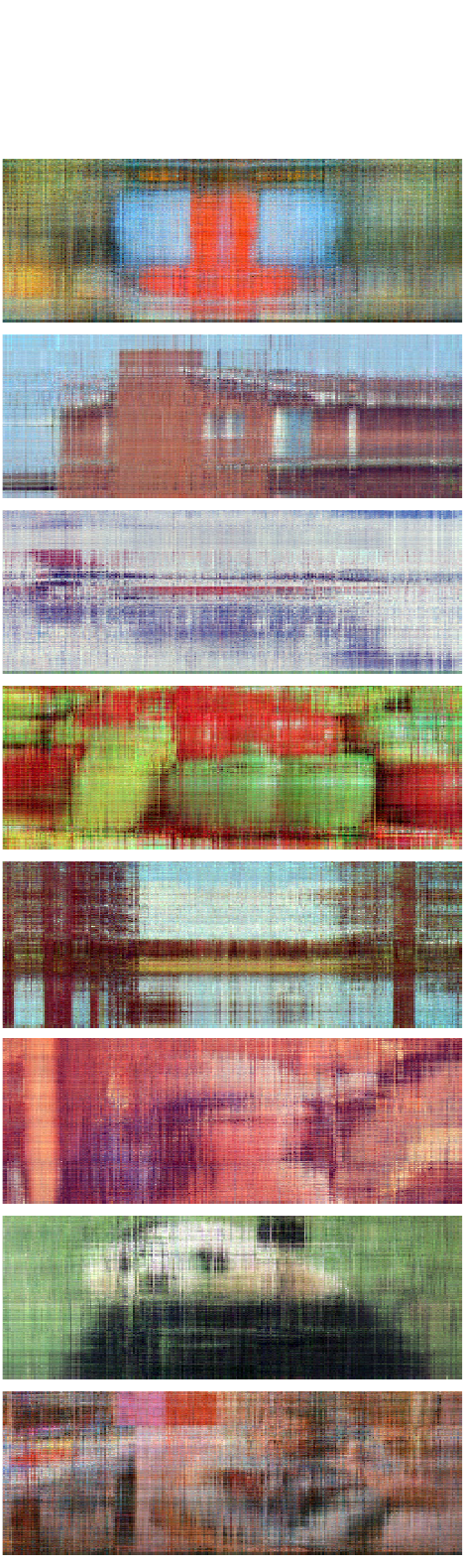}
	}
	\hspace{-0.18in}
	\subfigure[]{
		\includegraphics[width=2cm,height=13cm]{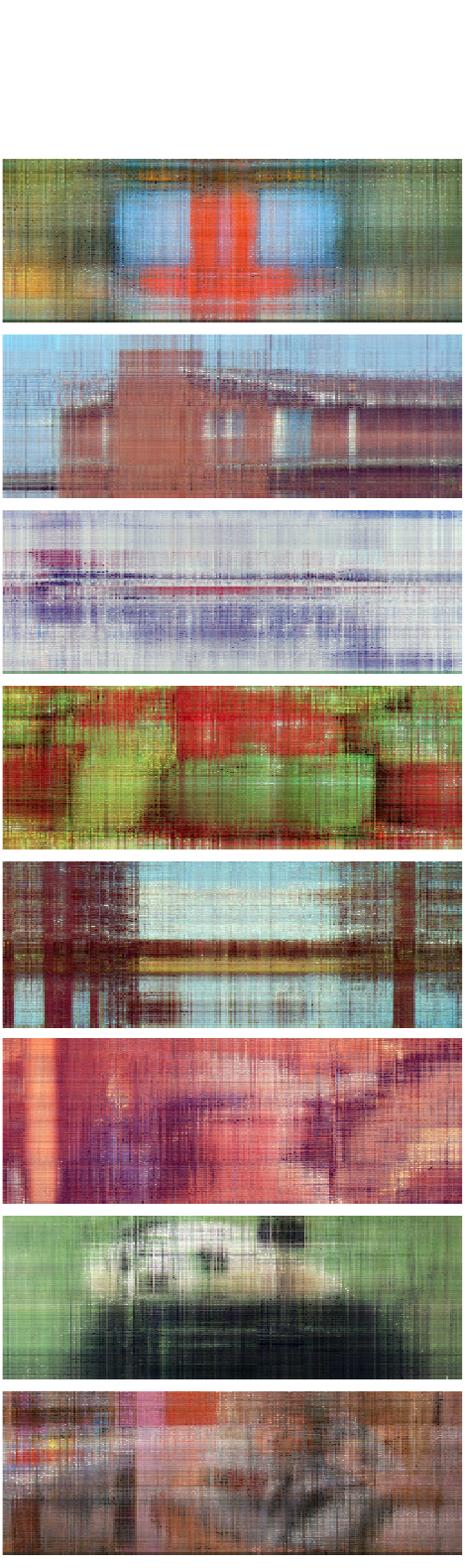}
	}
	\hspace{-0.18in}
	\subfigure[]{
		\includegraphics[width=2cm,height=13cm]{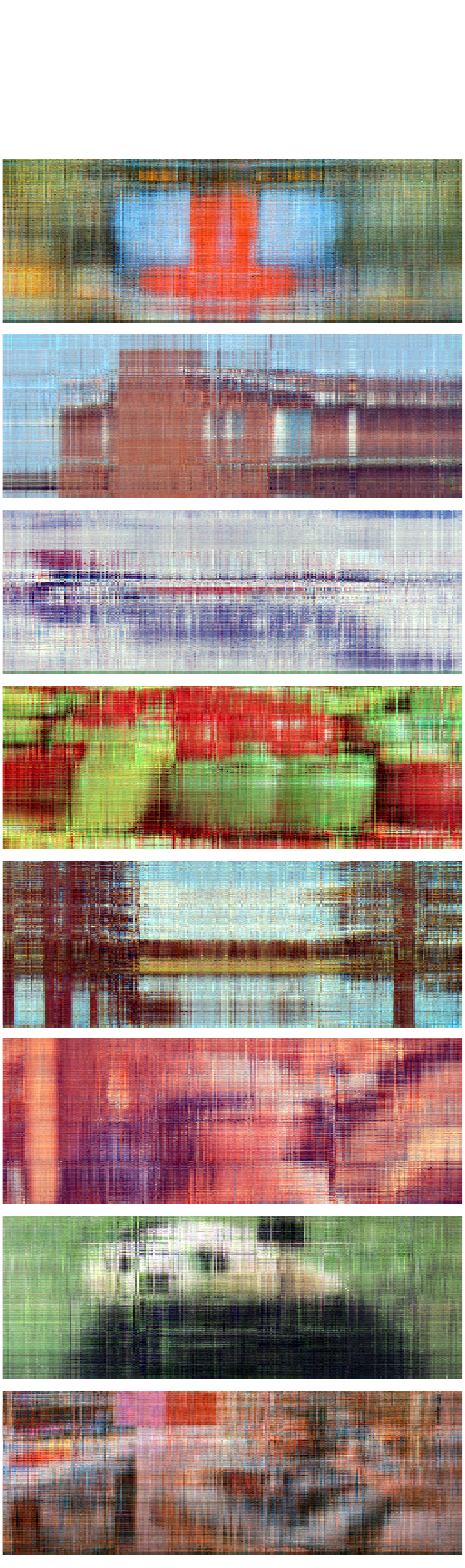}
	}
	\hspace{-0.18in}
	\subfigure[]{
		\includegraphics[width=2cm,height=13cm]{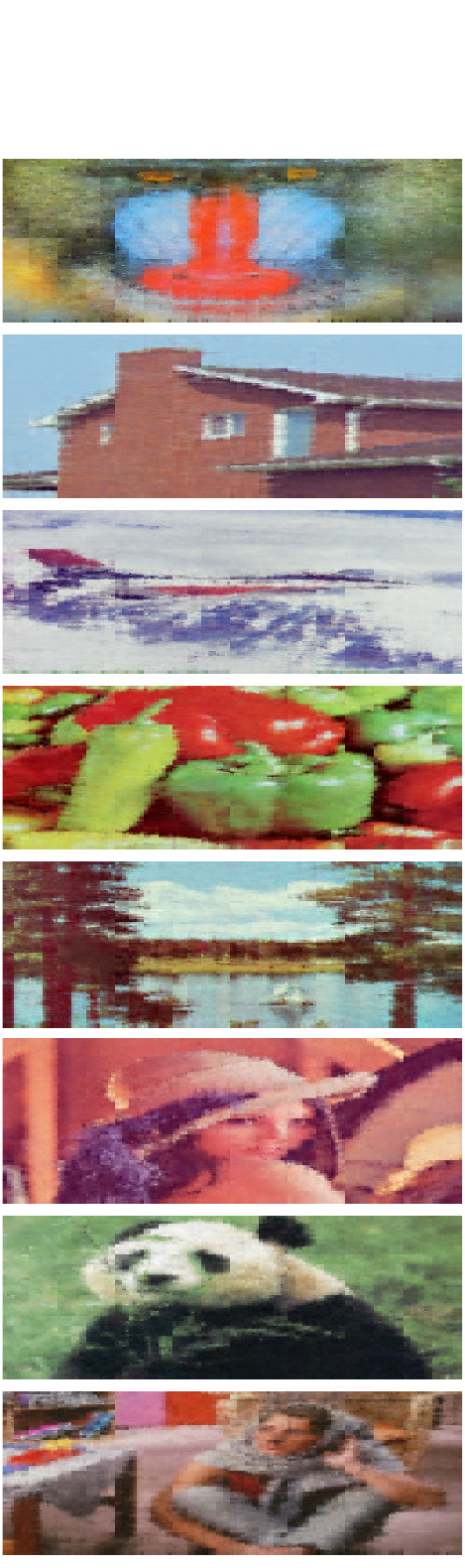}
	}
\caption{Recovered color images for random missing with ${\rm{SR}} = 10\%$. From left to right: the observed color images, the recovered results
	by t-SVD, SiLRTC-TT, TMac-TT, LRQA-2, LRQMC, TQLNA, and QTT-SRTD, respectively. From top to bottom: baboon, house, airplane, peppers, sailboat, lena, panda, and barbara.
}
\label{im_s1}
\end{figure*}
\begin{figure*}[htbp]
	\centering
	\subfigure[house]{\includegraphics[width=8.7cm,height=8cm]{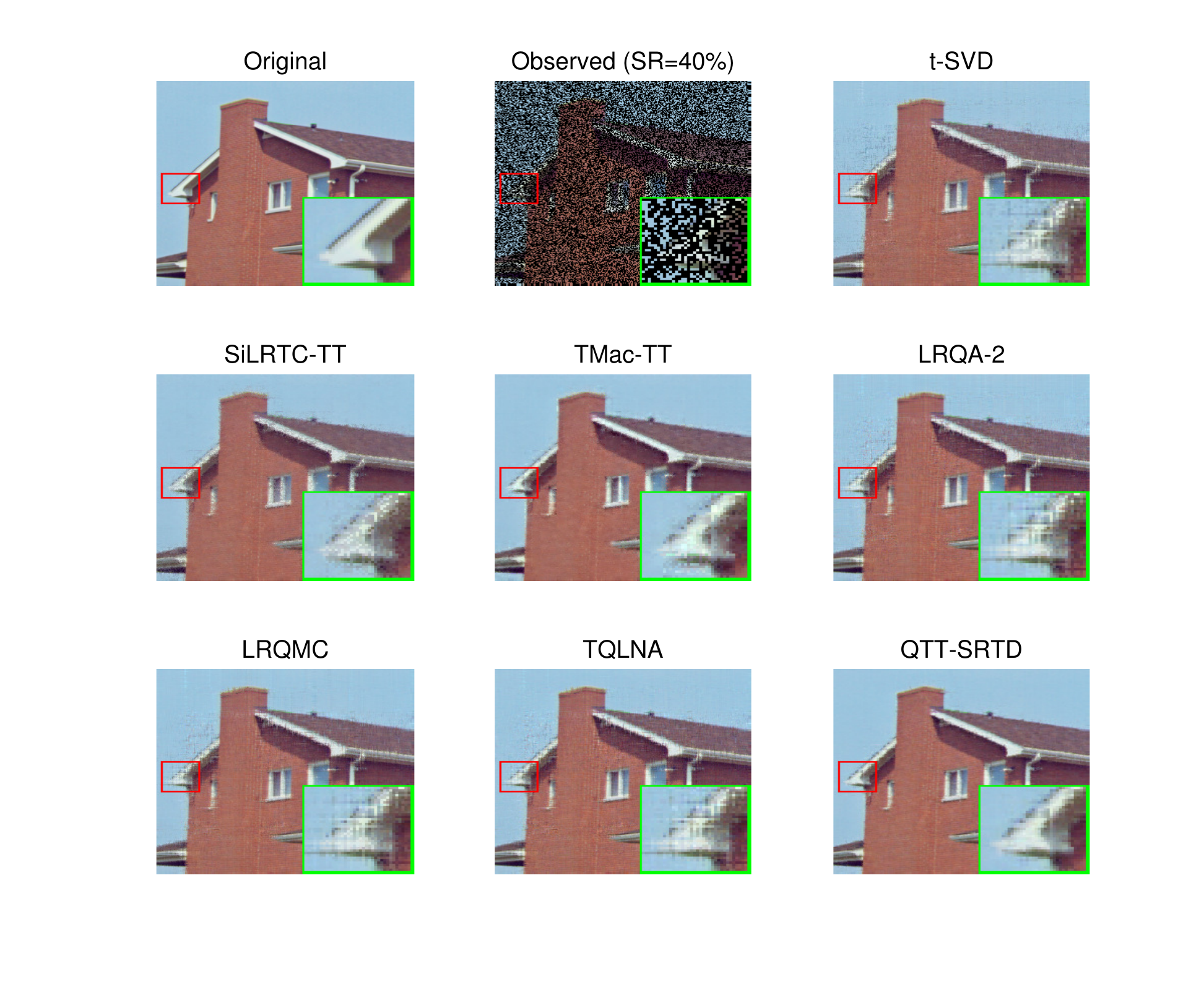}}\hspace{-1cm}
	\subfigure[lena]{\includegraphics[width=8.7cm,height=8cm]{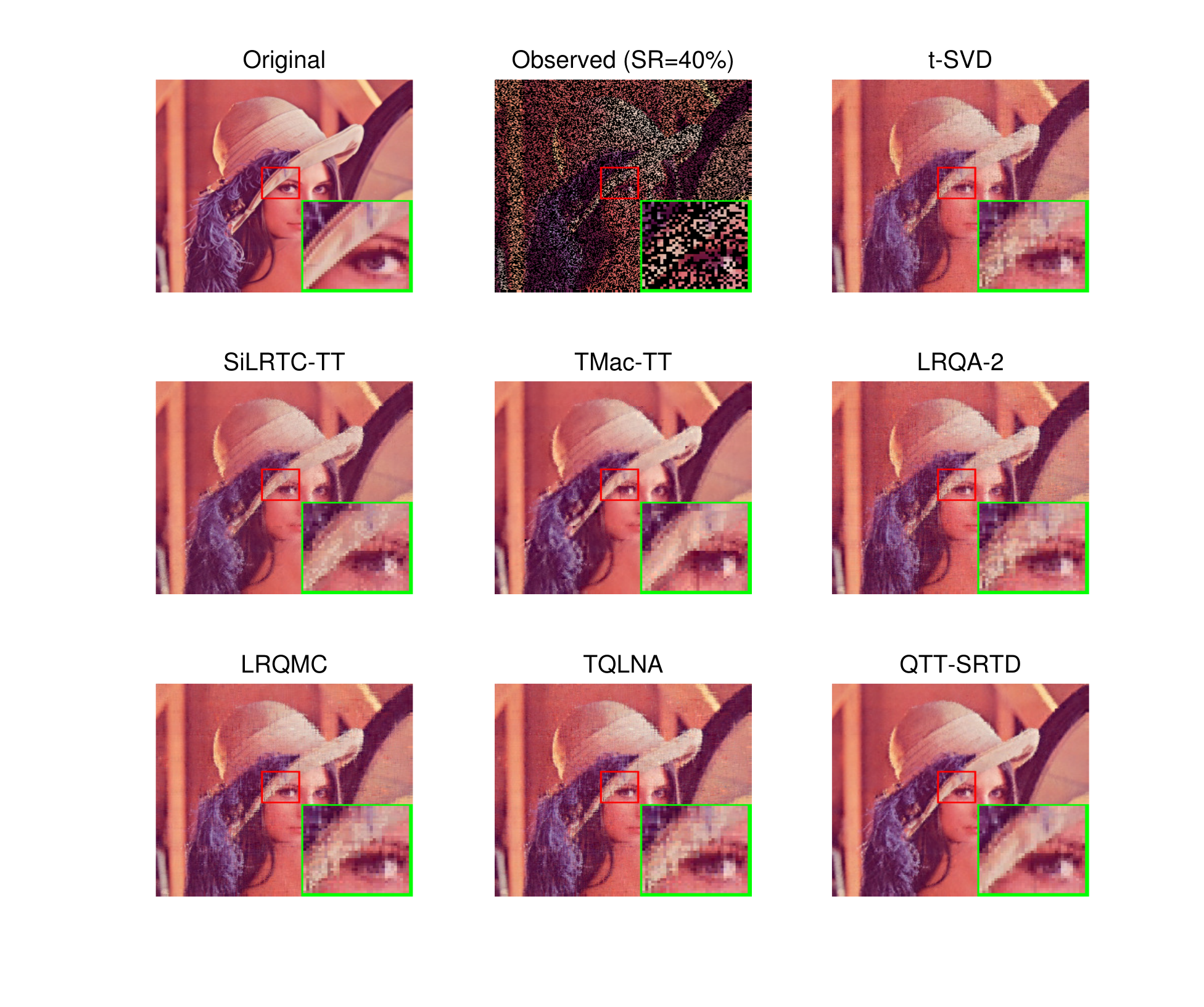}}
	\caption{Recovered two color images (house and lena) for random missing with ${\rm{SR}} = 40\%$.}
	\label{im_Es1}
\end{figure*}
\begin{table*}[htbp]
	\caption{The PSNR and SSIM values of different methods on the eight kinds of structural missing pixels (the format is PSNR/SSIM, and \textbf{bold} fonts denote the best performance).}
	\centering
	\resizebox{16.4cm}{2.2cm}{
		\begin{tabular}{|c|c|c|c|c|c|c|c|}		
			\hline
			Methods:&t-SVD \cite{DBLP:journals/tsp/ZhangA17} &SiLRTC-TT \cite{DBLP:journals/tip/BenguaPTD17}& TMac-TT \cite{DBLP:journals/tip/BenguaPTD17}& LRQA-2 \cite{DBLP:journals/tip/ChenXZ20}& LRQMC \cite{DBLP:journals/tip/MiaoK22} &TQLNA \cite{DBLP:journals/isci/YangMK22} &\textbf{QTT-SRTD} \\ \toprule
			\hline
			Images:  &PSNR/SSIM&PSNR/SSIM&PSNR/SSIM&PSNR/SSIM&PSNR/SSIM&PSNR/SSIM&PSNR/SSIM\\
			\hline
			baboon &29.848/0.963&30.459/0.968&30.268/0.967&29.258/0.954&29.829/0.962&29.426/0.959&\textbf{31.080}/\textbf{0.971}\\
			house&35.989/0.989&36.144/0.990&35.258/0.988&33.569/0.978&36.648/0.990&36.714/0.990 &\textbf{37.895}/\textbf{0.993}  \\
			airplane&30.317/0.952&32.067/0.972&31.069/0.972&29.630/0.913&30.455/0.956&30.860/0.958   &\textbf{33.115}/\textbf{0.980}  \\
			peppers&32.517/0.992&34.933/0.995&33.596/0.994&32.297/0.991&32.512/0.993&31.780/0.992 &\textbf{36.733}/\textbf{0.997} \\  
			sailboat&31.385/0.978&31.024/0.979&31.556/0.982&31.214/0.972&31.721/0.980&31.994/0.980&\textbf{33.744}/\textbf{0.988}  \\
			lena&33.399/0.992&35.634/0.995&33.591/0.993&31.699/0.987&33.474/0.993&33.445/0.993&\textbf{36.334}/\textbf{0.996} \\
			panda&30.111/0.943&31.766/0.961&32.228/0.962&29.898/0.921&30.888/0.950&30.905/0.947    &\textbf{32.954}/\textbf{0.968}\\
			barbara&30.763/0.971&31.518/0.977&30.614/0.973&30.730/0.957&30.957/0.971&31.531/0.971  &\textbf{33.377}/\textbf{0.980}\\\toprule
			\hline
			Aver. &31.791/0.973&32.943/0.980&32.273/0.979&31.037/0.959&32.061/0.974&32.082/0.974&\textbf{34.404}/\textbf{0.984}\\ \toprule
	\end{tabular}}
\label{table_2}
\end{table*}

Table \ref{table_1} lists the PSNR and SSIM values of different methods on the eight color images with four levels of sampling rates, and Figure \ref{aps_ssi_1} gives the average PSNR and SSIM values. It is observed that for different levels of sampling rates, the proposed QTT-SRTD  achieves the highest PSNR and SSIM values. Figure \ref{im_s1} and Figure \ref{im_Es1} visually display the recovered results by different methods for random missing with ${\rm{SR}} = 10\%$ and ${\rm{SR}} = 40\%$. From Figure \ref{im_s1} and Figure \ref{im_Es1}, one can see that the recovered results by the proposed QTT-SRTD are visually better than those
of the compared methods. From the zoom-in regions of recovered images, the proposed QTT-SRTD can
keep more details and smoothness of images compared with other methods. Experimental results of eight kinds of structural missing pixels are given in Table \ref{table_2} and Figure \ref{im_s2}, from which one can find that the proposed QTT-SRTD achieves the highest PSNR and SSIM values and also better visually.

\subsection{Color Video Inpainting}
In this experiment, we apply the proposed QTT-SRTD to color video inpainting from incomplete entries.

\textbf{Compared methods:} We compare the proposed QTT-SRTD with several well-known methods including t-SVD \cite{DBLP:journals/tsp/ZhangA17}, SiLRTC-TT \cite{DBLP:journals/tip/BenguaPTD17}, TMac-TT \cite{DBLP:journals/tip/BenguaPTD17}, and LRC-QT \cite{DBLP:journals/pr/MiaoKL20}.

\textbf{Parameter and initialization setting:} Same setting with color image inpainting.

\begin{figure*}[htbp]
	\centering
	\subfigure[]{
		\includegraphics[width=2.2cm,height=7cm]{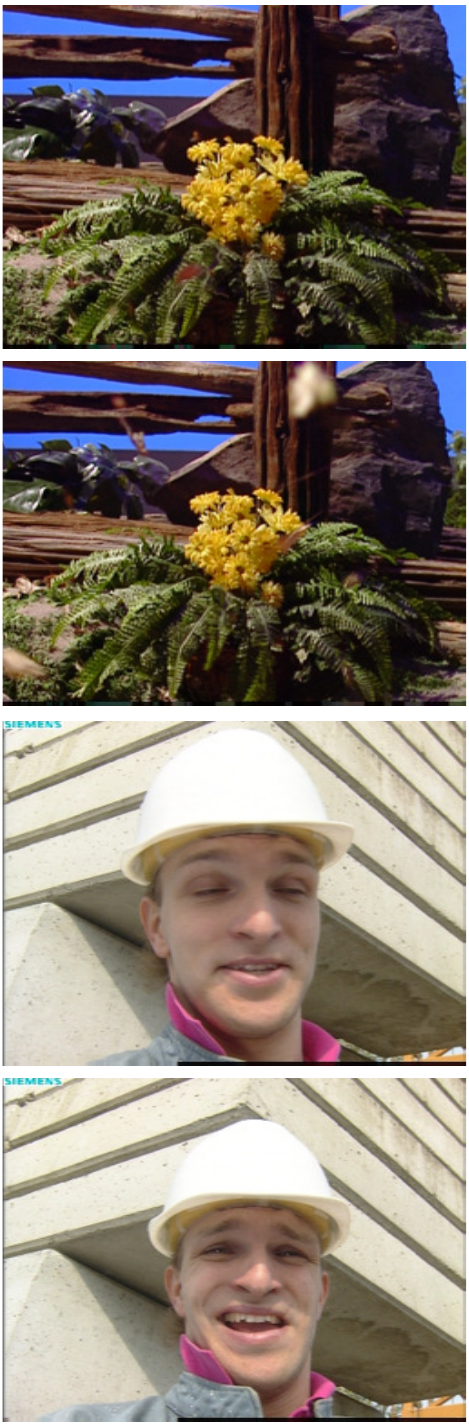}
	}
	\hspace{-0.18in}
	\subfigure[]{
		\includegraphics[width=2.2cm,height=7cm]{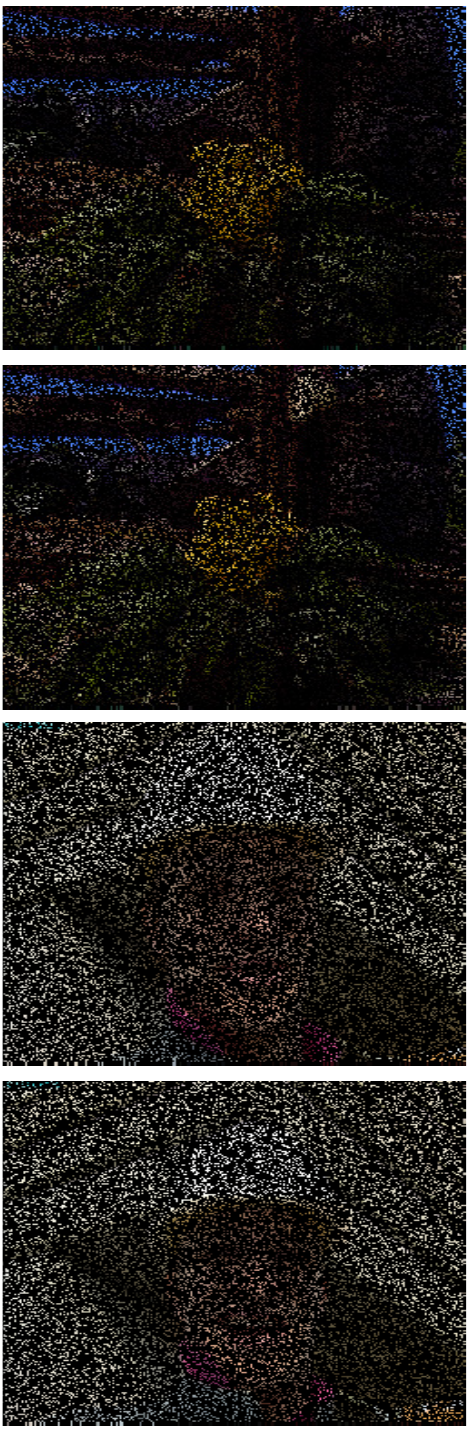}
	}
	\hspace{-0.18in}
	\subfigure[]{
		\includegraphics[width=2.2cm,height=7cm]{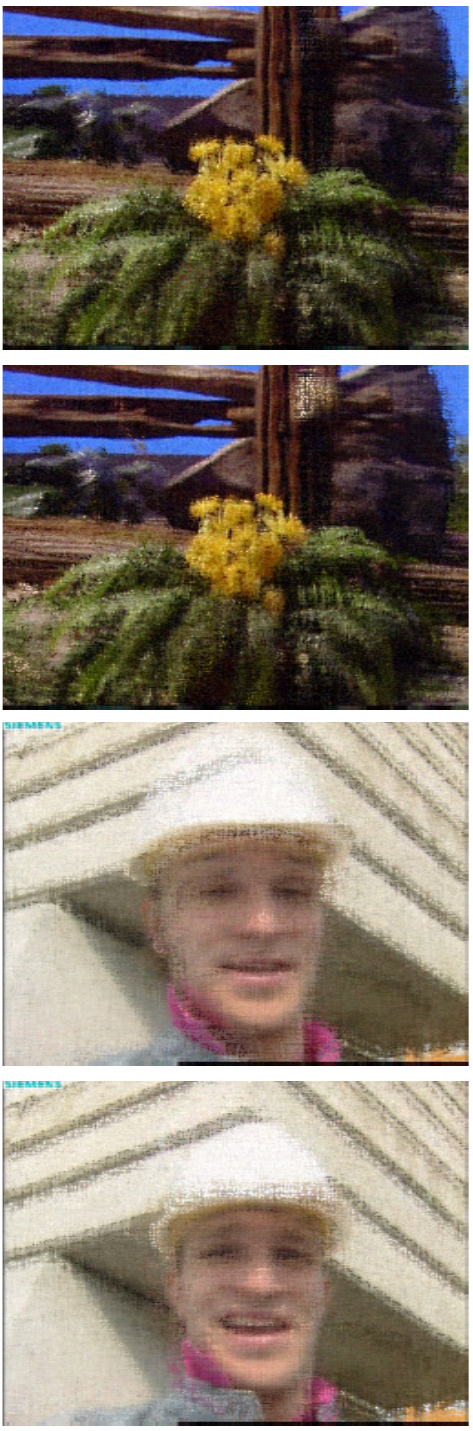}
	}
	\hspace{-0.18in}
	\subfigure[]{
		\includegraphics[width=2.2cm,height=7cm]{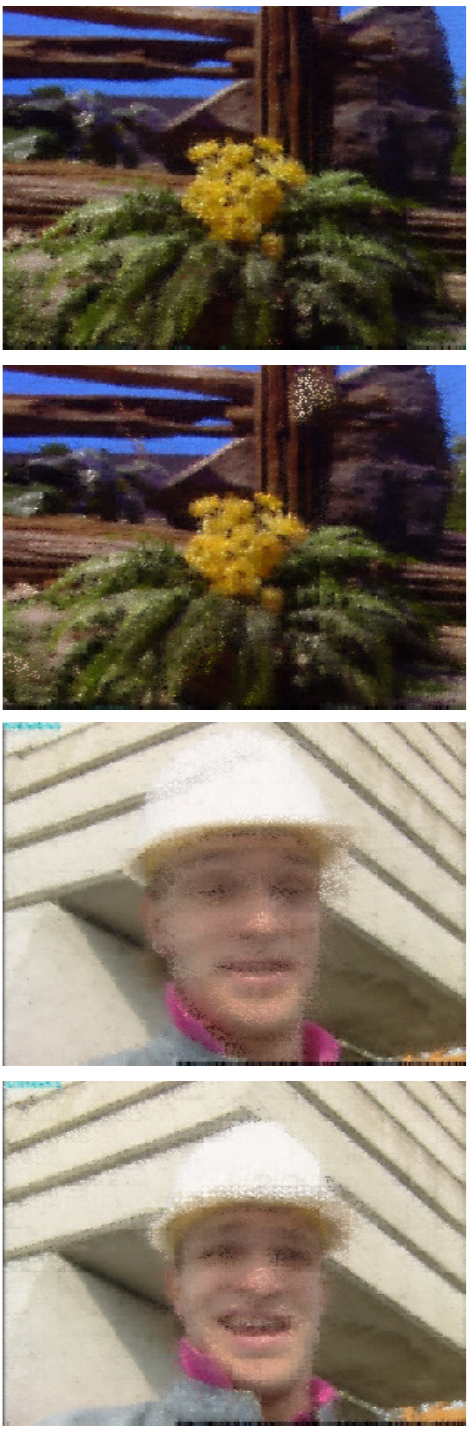}
	}
	\hspace{-0.18in}
	\subfigure[]{
		\includegraphics[width=2.2cm,height=7cm]{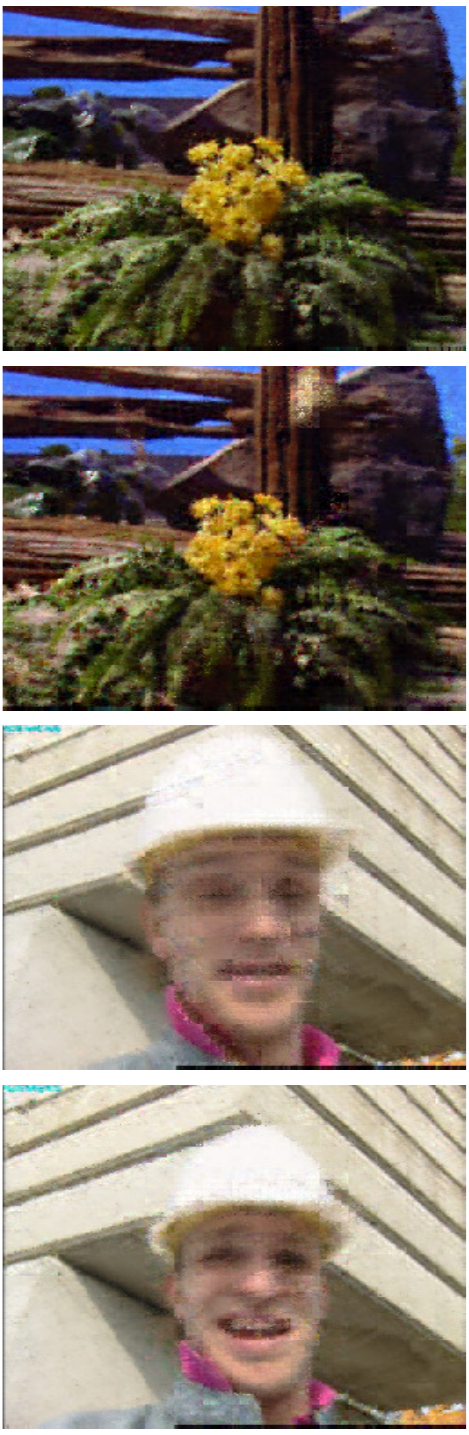}
	}
	\hspace{-0.18in}
	\subfigure[]{
		\includegraphics[width=2.2cm,height=7cm]{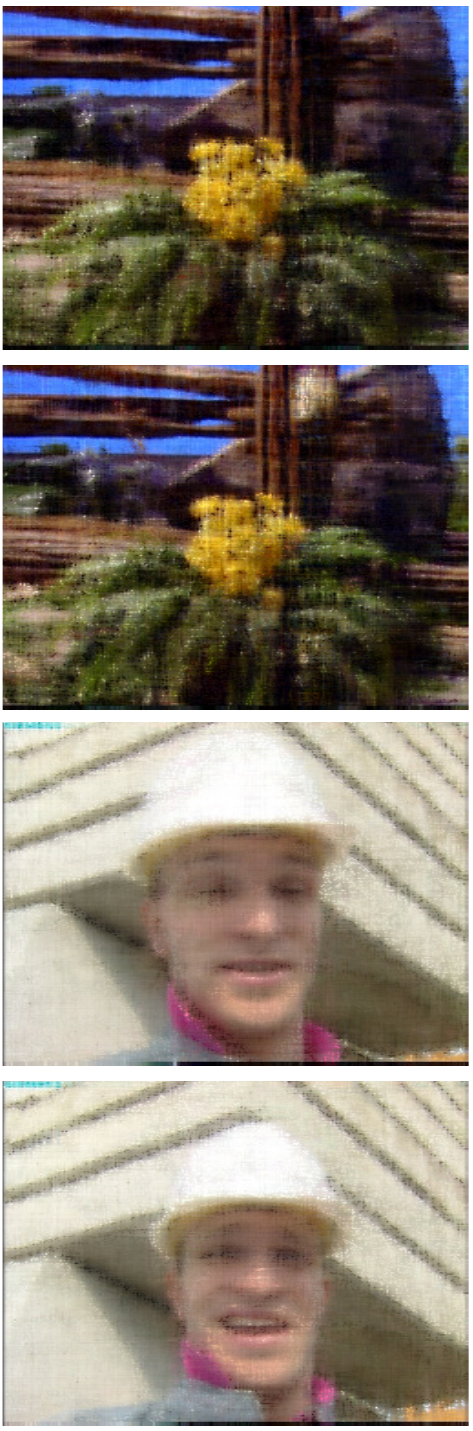}
	}
	\hspace{-0.18in}
	\subfigure[]{
		\includegraphics[width=2.2cm,height=7cm]{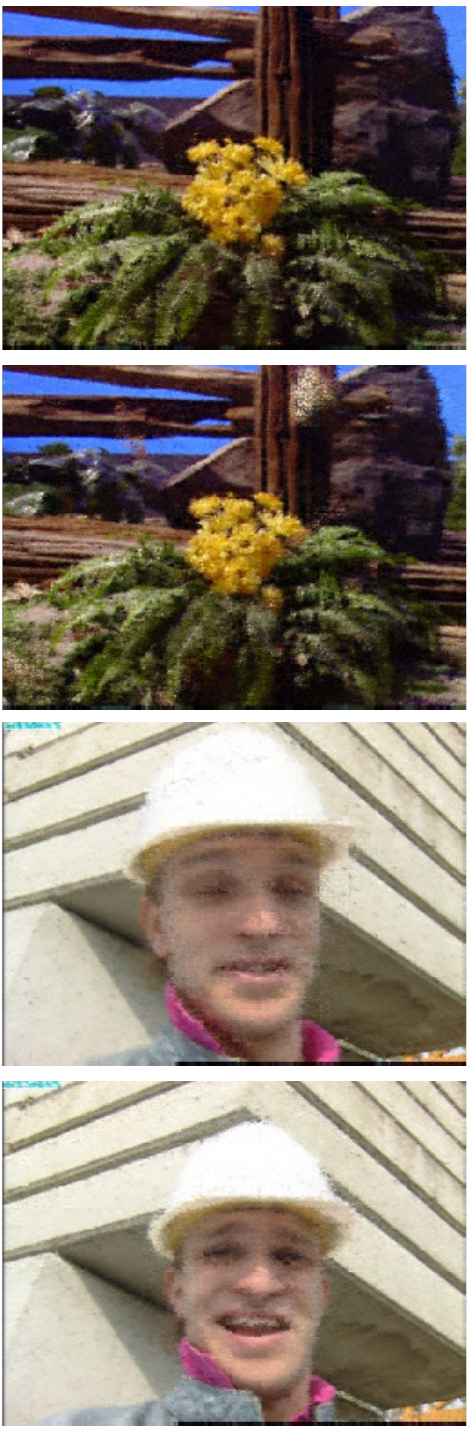}
	}
	\caption{Recovered two frames of testing color videos for random missing with ${\rm{SR}} = 20\%$. From left to right: the original, the observed, and the recovered results
		by t-SVD, SiLRTC-TT, TMac-TT, LRC-QT, and QTT-SRTD, respectively. From top to bottom: the $1$st and $20$th frames of ``Tempete'' and ``Foreman''. 
	}
	\label{vi_1}
\end{figure*}

\textbf{Test data and settings:} Numerical comparisons are implemented on two color videos\footnote{\url{\ http://trace.eas.asu.edu/yuv/}} (including ``Tempete'', and ``Foreman'', \emph{see} Figure \ref{vi_1}(a)) with size $256\times 256\times 20$. For the
QTT-SRTD, the color videos are transformed into fifth-order quaternion tensors with size $16\times 16\times 16\times 16\times 20$ by QKA. 
\begin{table*}[htbp]
	\caption{The PSNR and SSIM values (the average of $20$ frames) of different methods on the two color videos with two levels of sampling rates (the format is PSNR/SSIM, and \textbf{bold} fonts denote the best performance).}
	\centering
	\resizebox{12cm}{1.5cm}{
		\begin{tabular}{|c|c|c|c|c|c|}		
			\hline
			Methods:&t-SVD \cite{DBLP:journals/tsp/ZhangA17} &SiLRTC-TT \cite{DBLP:journals/tip/BenguaPTD17}& TMac-TT \cite{DBLP:journals/tip/BenguaPTD17}& LRC-QT \cite{DBLP:journals/pr/MiaoKL20} &\textbf{QTT-SRTD} \\ \toprule
			\hline
			Videos:  &\multicolumn{5}{c|}{${\rm{SR}}=10\%$}\\
			\hline
			Tempete &20.922/0.697&20.251/0.720&22.302/0.780&20.108/0.673&\textbf{23.249}/\textbf{0.821}\\
			Foreman&22.323/0.755&22.119/0.821&25.500/0.866&21.265/0.760&\textbf{27.266}/\textbf{0.899}  \\
			\hline
			\hline
			Videos:  &\multicolumn{5}{c|}{${\rm{SR}}=20\%$}\\
			\hline
			Tempete&23.229/0.802&22.425/0.820&23.683/0.836&21.842/0.747&\textbf{25.283}/\textbf{0.878}\\
			Foreman&25.020/0.840&24.993/0.891&26.850/0.901&24.053/0.843&\textbf{29.960}/\textbf{0.936}\\
			\hline		
	\end{tabular}}
	\label{table_v1}
\end{table*}

Figure \ref{vi_1} visually shows the recovered results of the $1$st and $20$th frames of ``Tempete'' and ``Foreman'' by different methods for random missing with ${\rm{SR}} = 20\%$. Table \ref{table_v1} gives the average PSNR and SSIM values of different methods on the two color videos with two levels of sampling rates ( ${\rm{SR}} = 10\%$ and  ${\rm{SR}} = 20\%$). We note that the proposed QTT-SRTD visually outperforms compared methods. The average PSNR and SSIM values achieved by the proposed QTT-SRTD are the highest.

\section{Conclusions}\label{sec_7}
In this paper, the QTT decomposition is first proposed, and based on it the QTT-rank is naturally defined. In addition, a general and flexible transform framework is defined for utilizing the local sparse prior of the quaternion tensor. Combining both the global low-QTT-rank and local sparse priors of the quaternion tensor, a novel LRQTC model, \emph{i.e.},  QTT-SRTD is proposed, which  is optimized by ADMM algorithm. Furthermore, The QKA is defined for QTT-rank minimization approaches to handle color images and better handle color videos. Extensive experimental results are given to show the effectiveness and superiority of the proposed method for inpainting problems of color images and color videos. In the
future, we would like to extend QTT-SRTD to other visual processing tasks, \emph{e.g.}, image denoising.

\appendices
\section{Quaternions}
\label{a_qsec1}
Quaternion space $\mathbb{H}$ was first introduced by W. Hamilton \cite{articleHamilton84} in 1843, which is an extension of the complex space $\mathbb{C}$. A quaternion $\dot{q}\in\mathbb{H}$ is given by 
\begin{equation}\small
	\label{aequ1}
	\dot{q}=q_{0}+q_{1}i+q_{2}j+q_{3}k,
\end{equation}
where $q_{l}\in\mathbb{R}\: (l=0,1,2,3)$, and $i, j, k$ are
imaginary units satisfying 
\begin{equation}\small
	\left\{
	\begin{array}{lc}
		i^{2}=j^{2}=k^{2}=ijk=-1,\\
		ij=-ji=k, jk=-kj=i, ki=-ik=j.
	\end{array}
	\right.
\end{equation}
If the real part $q_{0}=0$, $\dot{q}$ is called a pure quaternion. The addition and multiplication of quaternions are similar to complex numbers. It is worth noting that, unlike real and complex numbers, the multiplication of quaternions is not commutative, \emph{i.e.}, for two quaternions $\dot{p}\in\mathbb{H}$ and $\dot{q}\in\mathbb{H}$, in general, $\dot{p}\dot{q}\neq\dot{q}\dot{p}$.	
The conjugate and the modulus of a quaternion $\dot{q}$ are,
respectively, given by $\dot{q}^{\ast}=q_{0}-q_{1}i-q_{2}j-q_{3}k$ and $|\dot{q}|=\sqrt{\dot{q}\dot{q}^{\ast}}=\sqrt{q_{0}^{2}+q_{1}^{2}+q_{2}^{2}+q_{3}^{2}}$.

Analogously, a quaternion matrix $\dot{\mathbf{Q}}=(\dot{q}_{mn})\in\mathbb{H}^{M\times N}$ is given
by $\dot{\mathbf{Q}}=\mathbf{Q}_{0}+\mathbf{Q}_{1}i+\mathbf{Q}_{2}j+\mathbf{Q}_{3}k$, where $\mathbf{Q}_{l}\in\mathbb{R}^{M\times N}\: (l=0,1,2,3)$. A quaternion matrix $\dot{\mathbf{Q}}$ is called to be pure if $\mathbf{Q}_{0}=\mathbf{0}$. 
\begin{definition}\small
	(Cayley-Dickson form \cite{10029950538}): For any quaternion matrix $\dot{\mathbf{Q}}=\mathbf{Q}_{0}+\mathbf{Q}_{1}i+\mathbf{Q}_{2}j+\mathbf{Q}_{3}k\in\mathbb{H}^{M\times N}$, it can be represented as
	$\dot{\mathbf{Q}}=\mathbf{Z}_{1}+\mathbf{Z}_{2}j$,
	where $\mathbf{Z}_{1}=\mathbf{Q}_{0}+\mathbf{Q}_{1}i$ and $\mathbf{Z}_{2}=\mathbf{Q}_{2}+\mathbf{Q}_{3}i$ are complex matrices.
\end{definition}
\begin{theorem}\small
	(Quaternion singular value decomposition (QSVD) \cite{10029950538}): For any quaternion matrix $\dot{\mathbf{Q}}\in\mathbb{H}^{M\times N}$ with
	rank $r$, there exist unitary quaternion matrices $\dot{\mathbf{U}}\in\mathbb{H}^{M\times M}$ and $\dot{\mathbf{V}}\in\mathbb{H}^{N\times N}$ such that $\dot{\mathbf{Q}}=\dot{\mathbf{U}} \left[\begin{array}{cc}
		\mathbf{\Sigma}_{r}	&  \mathbf{0}\\
		\mathbf{0}	& \mathbf{0}
	\end{array}\right]\dot{\mathbf{V}}^{H}$, where  $\mathbf{\Sigma}_{r}={\rm{diag}}(\sigma_{1},\ldots,\sigma_{r})$ is a real diagonal matrix with $r$  positive entries $\sigma_{k}, \,(k=1,\ldots,r)$ on its diagonal, which are positive singular values of quaternion matrix $\dot{\mathbf{Q}}$.
\end{theorem}

The quaternion matrix Frobenius norm, $l_{1}$-norm, and quaternion nuclear norm are respectively given by
$\|\dot{\mathbf{Q}}\|_{F}=\sqrt{\sum_{m=1}^{M}\sum_{n=1}^{N}|\dot{q}_{mn}|^{2}}$,  $\|\dot{\mathbf{Q}}\|_{1}=\sum_{m=1}^{M}\sum_{n=1}^{N}|\dot{q}_{mn}|$, and $\|\dot{\mathbf{Q}}\|_{\ast}=\sum_{k}\sigma_{k}$.	

\begin{definition}(Quaternion tensor \cite{DBLP:journals/pr/MiaoKL20}) A multidimensional array or an $N$th-order tensor is named a quaternion tensor if its entries are quaternion numbers, i.e., $\dot{\mathcal{X}}=(\dot{x}_{i_{1}i_{2}\ldots i_{N}})\in\mathbb{H}^{I_{1}\times I_{2} \times\ldots \times I_{N}}
		=\mathcal{X}_{0}+\mathcal{X}_{1}i+\mathcal{X}_{2}j+\mathcal{X}_{3}k$, where $\mathcal{X}_{l}\in\mathbb{R}^{I_{1}\times I_{2} \times\ldots \times I_{N}}\: (l=0,1,2,3)$, $\dot{\mathcal{X}}$ is pure if $\mathcal{X}_{0}$ is a zero tensor.
\end{definition}
\begin{definition}\label{nuft}(Mode-$n$ unfolding \cite{DBLP:journals/pr/MiaoKL20}) For an $N$th-order quaternion tensor $\dot{\mathcal{X}}\in\mathbb{H}^{I_{1}\times I_{2} \times\ldots \times I_{N}}$, its mode-$n$ unfolding is denoted by $\dot{\mathbf{X}}_{(n)}$ and arranges the mode-$n$ fibers to be the columns of the quaternion matrix, i.e.,
${\rm{unfold}}_{(n)}(\dot{\mathcal{X}})=\dot{\mathbf{X}}_{(n)}\in\mathbb{H}^{I_{n}\times I_{1}\ldots I_{n-1}I_{n+1}\ldots I_{N}}$. 
	The $(i_{1},i_{2},\ldots,i_{N})$th entry of $\dot{\mathcal{X}}$ maps to the $(i_{n},x)$th entry of $\dot{\mathbf{X}}_{(n)}$, where $x=1+\sum_{\substack{p=1\\p\neq n}}^N(i_{p}-1)X_{p}$	with $X_{p}=\prod_{\substack{d=1\\d\neq n}}^{p-1}I_{d}$.
\end{definition}
\begin{definition}
	\label{mode product}
	(The $n$-mode product \cite{DBLP:journals/corr/abs-2101-00364}) The $n$-mode product of a quaternion tensor $\dot{\mathcal{X}}\in\mathbb{H}^{I_{1}\times I_{2} \times\ldots \times I_{N}}$ with a quaternion matrix $\dot{\mathbf{U}}\in\mathbb{H}^{M\times I_{n}}$ is denoted by$
\dot{\mathcal{Y}}=\dot{\mathcal{X}}\times_{n}\dot{\mathbf{U}}\in\mathbb{H}^{I_{1}\times  \ldots \times I_{n-1} \times M \times I_{n+1}\times \ldots \times I_{N}}$ with entries $\dot{y}_{i_{1}\ldots i_{n-1} m i_{n+1}\ldots  i_{N}}=\sum_{i_{n}=1}^{I_{n}}\dot{u}_{mi_{n}}\dot{x}_{i_{1}i_{2}\ldots i_{N}}$.
\end{definition}

One can find more details on quaternions in \cite{10029950538, Girard2007Quaternions, DBLP:journals/tsp/MiaoK20,DBLP:journals/pr/MiaoKL20,DBLP:journals/corr/abs-2101-00364}.
	
\section{The proof of Theorem \ref{theqtt}}
\label{a_sec1}
\begin{proof}
$\dot{\mathcal{X}}$ with a QTT decomposition (\ref{eqqtt}) can be represented as 
\begin{equation}\label{equa1}\small	
		\dot{x}_{i_{1},i_{2},\ldots i_{N}}=\sum_{l_{n}=1}^{r_{n}}\dot{\mathcal{P}}(i_{1},\ldots i_{n},l_{n})\dot{\mathcal{Q}}(l_{n},i_{n+1},\ldots i_{N}),
\end{equation}
where $\dot{\mathcal{P}}\in \mathbb{H}^{I_{1}\times \ldots \times I_{n}\times r_{n}}$ and $\dot{\mathcal{Q}}\in \mathbb{H}^{r_{n}\times I_{n+1}\times \ldots \times I_{N}}$, their entries are calculated by
\begin{equation*}\small
\begin{split}
\dot{\mathcal{P}}(i_{1},\ldots i_{n},l_{n})=	\sum_{l_{1},\ldots,l_{n-1}=1}^{r_{1},\ldots,r_{n-1}}&\dot{\mathcal{G}}_{1}(1,i_{1},l_{1})\dot{\mathcal{G}}_{2}(l_{1},i_{2},l_{2})\ldots\\
&\dot{\mathcal{G}}_{n}(l_{n-1},i_{n},l_{n})
\end{split}
\end{equation*}
\begin{equation*}\small
	\begin{split}
		\dot{\mathcal{Q}}(l_{n},i_{n+1},\ldots, i_{N})=\!\!\!\!	\sum_{l_{n+1},\ldots,l_{N-1}=1}^{r_{n+1},\ldots,r_{N-1}}\!\!\!\!&\dot{\mathcal{G}}_{n+1}(l_{n},i_{n+1},l_{n+1})\ldots\\ &\dot{\mathcal{G}}_{N}(l_{N-1},i_{N},1).		
	\end{split}
\end{equation*}
Equation (\ref{equa1}) is equivalent to the following one:
\begin{equation*}\small
\dot{\mathbf{X}}_{[n]}=\dot{\mathbf{P}}_{[n]}\dot{\mathbf{Q}}_{[1]},	
\end{equation*}
where $\dot{\mathbf{P}}_{[n]}\in \mathbb{H}^{\Pi_{j=1}^{n}I_{j}\times r_{n}}$ and $\dot{\mathbf{Q}}_{[1]}\in \mathbb{H}^{r_{n}\times \Pi_{j=n+1}^{N}I_{j}}$. Hence, we have
\begin{equation*}\small
{\rm{rank}}(\dot{\mathbf{X}}_{[n]})={\rm{rank}}(\dot{\mathbf{P}}_{[n]}\dot{\mathbf{Q}}_{[1]}) \leq r_{n}.
\end{equation*} 
Equality can be achieved following the recursive procedure leading to (\ref{eqqtt2}) with
minimal $(r_{1},r_{2},\ldots, r_{N-1})$ in every step described in Remark 1. Specifically, the first step (\ref{equde1}) is clearly a rank revealing decomposition of $\dot{\mathbf{X}}_{[1]}$ so the rank of that
quaternion matrix can be used as $r_{1}$. The minimal admissible $r_{2}$ in (\ref{equde2})  is
the rank of $\dot{\mathbf{H}}_{1{[2]}}$. Next, we show that ${\rm{rank}}(\dot{\mathbf{H}}_{1{[2]}})\leq{\rm{rank}}(\dot{\mathbf{X}}_{[2]})$. Assume $\dot{\mathbf{z}}\in\mathbb{H}^{\Pi_{j=3}^{N}I_{j}\times 1}$ such that $\dot{\mathbf{X}}_{[2]}\dot{\mathbf{z}}=0$. Let $\dot{\mathbf{y}}=\dot{\mathbf{H}}_{1{[2]}}\dot{\mathbf{z}}$, since $\dot{\mathcal{G}}_{1}(1,:,:)$ has rank $r_{1}$, then (\ref{equde1}) yields $0=\sum_{l_{1}=1}^{r_{1}}\dot{\mathcal{G}}_{1}(1,i_{1},l_{1})\dot{\mathbf{y}}(l_{1})$. Thus, $\dot{\mathbf{y}}=0$, which implies that ${\rm{rank}}(\dot{\mathbf{H}}_{1{[2]}})\leq{\rm{rank}}(\dot{\mathbf{X}}_{[2]})$ indeed holds. Hence, $r_{2}={\rm{rank}}(\dot{\mathbf{X}}_{[2]})$ by (\ref{equin}). For the subsequent $r_{3},\cdots,r_{N-1}$, one can proceed by using a similar argument.
\end{proof}

\section{A general way to generate quaternion transforms from the corresponding traditional transforms} \label{a_sec2}
In \cite{feng2008quaternion}, the authors presented the extension of DCT to the quaternion domain and showed how to calculate the QDCT of a quaternion matrix using its Cayley-Dickson form. In this section, we extend the calculation method of QDCT in \cite{feng2008quaternion} to any appropriate quaternion transforms, \emph{i.e.}, a general way to generate quaternion transforms from the corresponding traditional transforms.

Suppose $\mathcal{CT}(\cdot)$ is a traditional common transform, then the corresponding quaternion transform $\mathcal{QT}(\dot{\mathbf{A}})$ for a given quaternion matrix is obtained by the following procedure:
\begin{enumerate}
	\item For a given quaternion matrix $\dot{\mathbf{A}}\in\mathbb{H}^{M\times N}$, transform it into
	the Cayley-Dickson form such that
	\begin{equation*}\small
		\dot{\mathbf{A}}=\mathbf{C}+\mathbf{D}j,
	\end{equation*}
where $\mathbf{C}\in\mathbb{C}^{M\times N}$ and $\mathbf{D}\in\mathbb{C}^{M\times N}$ are complex matrices with the same size as $\dot{\mathbf{A}}$. 
    \item Calculate $\mathcal{CT}(\mathbf{C})$ and $\mathcal{CT}(\mathbf{D})$ respectively, then using them to form a full quaternion matrix, \emph{i.e.},
    \begin{equation*}\small
    	\dot{\tilde{\mathbf{A}}}=\mathcal{CT}(\mathbf{C})+\mathcal{CT}(\mathbf{D})j.
    \end{equation*}
   \item Multiply $\dot{\tilde{\mathbf{A}}}$ with a pure quaternion factor $\dot{\mu}$ (which satisfies that $\dot{\mu}^{2}=-1$) to obtain the final result, \emph{i.e.}, 
   \begin{equation}\label{qtransL}\small
   		\mathcal{QT}_{L}(\dot{\mathbf{A}}):=\dot{\mu}\dot{\tilde{\mathbf{A}}}=\dot{\mu}(\mathcal{CT}(\mathbf{C})+\mathcal{CT}(\mathbf{D})j)
   \end{equation}
or
   \begin{equation}\label{qtransR}\small
	\mathcal{QT}_{R}(\dot{\mathbf{A}}):=\dot{\tilde{\mathbf{A}}}\dot{\mu}=(\mathcal{CT}(\mathbf{C})+\mathcal{CT}(\mathbf{D})j)\dot{\mu}.  	
\end{equation}
Note that due to the non-commutative multiplication rule of quaternions, the form of $\mathcal{QT}(\dot{\mathbf{A}})$ has two categories, left-handed form (\ref{qtransL}) (default form in the paper) and right-handed form (\ref{qtransR}). Besides, the corresponding inverse transforms can be easily obtained by following similar steps.
\end{enumerate}

\bibliographystyle{IEEEtran}
\bibliography{Myreference}

\end{document}